\documentclass[12pt,a4paper]{article}

\usepackage{amsmath,amsfonts,latexsym,amssymb}
\usepackage{verbatim,amsthm,curves,graphics}
\usepackage{mathrsfs}

\theoremstyle{definition}
\newtheorem{thm}{Theorem}[section]

\newtheorem{lemma}{Lemma}[section]
\newtheorem{prop}{Proposition}[section]

\newtheorem{defn}{Definition}[section]

\begin{document}

\renewcommand{\j}{{\mathcal R}}
\newcommand{\xsim}{\stackrel{x}{\sim}}
\newcommand{\xequiv}{\stackrel{x,\mu}{\sim}}
\newcommand{\chisub}{\stackrel{\chi}{\subset}}
\newcommand{\myI}{\stackrel{\circ}{I}}
\newcommand{\mr}{\mathbb{R}} 
\newcommand{\mi}{\mathbb{J}} 
\newcommand{\mz}{\mathbb{Z}} 
\newcommand{\mc}{\mathbb{C}} 
\newcommand{\mh}{\mathbb{H}} 
\newcommand{\ms}{\mathbb{S}} 
\newcommand{\mn}{\mathbb{N}} 
\newcommand{\F}{{\bf F}} 
\newcommand{\beps}{\mbox{\boldmath $\epsilon$}} 
\newcommand{\pgator}{/\!\!\! S}
\newcommand{\rP}{\operatorname{P}}
\newcommand{\e}{\operatorname{e}}
\newcommand{\car}{\operatorname{CAR}}
\newcommand{\rSpin}{\operatorname{Spin}}  
\newcommand{\rSO}{\operatorname{SO}}      
\newcommand{\rO}{\operatorname{O}}        
\newcommand{\rCliff}{\operatorname{Cliff}}
\newcommand{\WF}{\operatorname{WF}}       
\newcommand{\Pol}{{\rm WF}_{pol}}         
\newcommand{\cC}{\mathscr{C}}             
\newcommand{\cA}{\mathcal{A}}                
\newcommand{\cW}{w}                
\newcommand{\cU}{\mathcal{U}}                
\newcommand{\cO}{\mathcal{O}}
\newcommand{\bO}{\mbox{\boldmath $O$}}
\newcommand{\Star}{\operatorname{\mbox{\huge $\star$}}}
\newcommand{\M}{\mathcal{M}}             
\newcommand{\N}{\mathcal{N}}             
\newcommand{\cS}{\mathcal{S}}             
\newcommand{\cF}{\mathcal{F}}             
\newcommand{\cE}{\mathcal{E}}             
\newcommand{\cH}{\mathcal{H}}
\newcommand{\bW}{\mbox{\boldmath $W$}}
\newcommand{\bw}{\mbox{\boldmath $w$}}
\newcommand{\cP}{\mathcal{P}}
\newcommand{\cT}{\mathscr{T}}
\newcommand{\cN}{\mathscr{N}}
\newcommand{\cD}{\mathcal{D}}             
\newcommand{\cL}{{\mathcal{L}^\uparrow_+}}             
\newcommand{\cQ}{\mathscr{Q}}             
\newcommand{\cJ}{\mathcal{J}}             
\newcommand{\bS}{\mbox{\boldmath $S$}}
\newcommand{\bF}{\mbox{\boldmath $F$}}
\newcommand{\bU}{\mathcal{U}}
\newcommand{\cV}{\mathcal{V}}
\newcommand{\glob}{{\textrm{\small global}}}
\newcommand{\loca}{{\textrm{\small local}}}
\newcommand{\singsupp}{\operatorname{singsupp}}
\newcommand{\dom}{\operatorname{dom}}
\newcommand{\clo}{\operatorname{clo}}
\newcommand{\supp}{\operatorname{supp}}
\newcommand{\rd}{{\rm d}}                 
\newcommand{\mslash}{/\!\!\!}             
\newcommand{\slom}{/\!\!\!\omega}         
\newcommand{\dirac}{/\!\!\!\nabla}        
\newcommand{\myid}{\leavevmode\hbox{\rm\small1\kern-3.8pt\normalsize1}}
\newcommand{\bv}{\operatorname*{B.V.}}
\newcommand{\ran}{\operatorname{ran}}
\newcommand{\sd}{{\rm sd}}
\newcommand{\reg}{\,{\rm l.n.o.}}
\newcommand{\isom}{\iota}
\newcommand{\wk}{{\bf k}}
\newcommand{\ws}{{\bf s}}
\newcommand{\lno}{:\!}
\newcommand{\rno}{\!:}
\newcommand{\clim}{\operatorname*{coin.lim}}
\newcommand{\mydef}{\stackrel{\textrm{def}}{=}}
\renewcommand{\min}{{\textrm{\small int}\,\mathscr{L}}}
\renewcommand{\max}{\operatorname{max}}
\renewcommand{\Im}{\operatorname{Im}}
\renewcommand{\Re}{\operatorname{Re}}
\newcommand{\Exp}{\operatorname{Exp}}
\newcommand{\Ad}{\operatorname{Ad}}
\newcommand{\sa}{\mathfrak{sa} } 
\newcommand{\so}{\mathfrak{so} } 
\renewcommand{\o}{\mathfrak{o} } 
\newcommand{\su}{\mathfrak{su} } 
\newcommand{\sq}{\mathfrak{sq} } 
\renewcommand{\sp}{\mathfrak{sp} } 
\newcommand{\gl}{\mathfrak{gl}} 
\newcommand{\g}{g_{ab}}	
\newcommand{\bfeta}{\mbox{\boldmath $\eta$}}	
\newcommand{\h}{\mathfrak{h} } 
\newcommand{\f}{\mathfrak{f} } 
\newcommand{\p}{\mathfrak{p} } 
\newcommand{\U}{\operatorname{U}} 
\newcommand{\Z}{\operatorname{Z}} 
\newcommand{\SO}{\operatorname{SO} } 
\newcommand{\SU}{\operatorname{SU} } 
\renewcommand{\O}{\operatorname{O} }
\newcommand{\SP}{\operatorname{Sp} } 
\newcommand{\SL}{\operatorname{SL}} 
\newcommand{\G}{\mathcal{G}} 
\newcommand{\Der}{\operatorname{Der}} 
\newcommand{\Str}{\operatorname{Str}} 
\newcommand{\End}{\operatorname{End}} 
\newcommand{\Cl}{\operatorname{Cl}} 
\newcommand{\ds}{\dot{+}}
\renewcommand{\aa}{\alpha}
\renewcommand{\ln}{\log}
\renewcommand{\d}{\delta }
\newcommand{\x}{\bar{x} } 
\newcommand{\y}{\bar{y} }
\newcommand{\s}{\bar{s} }
\newcommand{\diag}{\operatorname{diag}}
\renewcommand{\Ad}{\operatorname{Ad}}
\newcommand{\sgn}{\operatorname{sgn}}
\newcommand{\tr}{\operatorname{tr}} 
\newcommand{\per}{\operatorname{per}} 
\newcommand{\Tri}{\operatorname{Tri}}
\renewcommand{\square}{\nabla^a \nabla_a}

\title{A General PCT Theorem for the Operator Product Expansion in Curved Spacetime}
\author{Stefan Hollands\thanks{Electronic mail: \tt stefan@bert.uchicago.edu}\\
       \it{Enrico Fermi Institute, Department of Physics,}\\
       \it{University of Chicago, 5640 Ellis Ave.,}\\
       \it{Chicago IL 60637, USA}
        }
\date{\today}

\maketitle

\begin{abstract}
We consider the operator product expansion for quantum field theories on general 
analytic 4-dimensional curved spacetimes within an axiomatic framework. We prove 
under certain general, model-independent assumptions that such an expansion 
necessarily has to be invariant under 
a simultaneous reversal of parity, time, and charge (PCT) in the following sense: 
The coefficients in the expansion of a product of fields on 
a curved spacetime with a given choice of time and space orientation are equal 
(modulo complex conjugation) to the coefficients for the product of the corresponding 
charge conjugate fields on the spacetime with the opposite time and space orientation. 
We propose that this result should be viewed as a replacement of the usual PCT theorem in 
Minkowski spacetime, at least in as far as the algebraic structure of the quantum fields at 
short distances is concerned.
\end{abstract}

\pagebreak

\section{Introduction}

The operator product expansion~\cite{w,z} 
states that the product of any finite number of field operators localized at nearby 
points can be approximated by a sum of products of  
c-number coefficient functions of the coordinates of the points relative to 
a reference point, times fields that are localized at the reference point. Furthermore, 
of these coefficient functions, only finitely many are singular as the spacetime points approach
the reference point. In mathematical symbols, 
if the fields in the theory are denoted by the generic symbol $\phi^{(i)}$ 
(with $(i)$ a label that distinguishes the various kinds of fields), then the operator expansion 
is an expansion of the form
\begin{equation}
\label{ope}
\phi^{(1)}(y_1) \phi^{(2)}(y_2) \cdots \phi^{(n)}(y_n) \sim \sum_{(i)} c^{(i)}(y_1, y_2, \dots, y_n)
\phi^{(i)}(x). 
\end{equation}
The notation ``$\sim$'' means that the difference between the expectation value in any ``reasonable'' state of the 
left side and the expectation value of a suitable finite partial sum on right side goes to zero as the points 
$y_1, \dots, y_n$ approach the reference point\footnote{One can of course take advantage of translation 
invariance in Minkowksi spacetime to set the reference point $x$ to the origin in Minkowkski space, which 
is done in usual formualtions of the operator product expansion. We are avoiding this since it 
has no invariant meaning on a curved spacetime.}, $x$. Moreover, 
the rate at which this difference goes to zero can be made arbitrary by including sufficiently many terms
in the partial sum. In practice, the operator product expansion is useful to find approximate expressions at short distances (high momenta) 
for the expectation value of a product of $n$ fields when the corresponding expectation values of the 
singly localized fields $\phi^{(i)}$ on the right side of eq.~\eqref{ope} are known, for example experimentally, 
and where the coefficients $c^{(i)}$ can be calculated. Such techniques have been used successfully e.g. to 
gain insights into the internal structure of hadrons.

In Minkowski spacetime, model-independent derivations of the operator product expansion 
from first principles within an axiomatic framework, including a precise specification of the nature of 
states for which it holds, have been given in various contexts~\cite{wz,fj,ss,l,b}. The derivation that is in our 
opinion most general and physically best motivated seems to be that of~\cite{b}, 
(which is based in turn on earlier work by~\cite{fh,bw}) and our analysis builds partly 
on the results and ideas of this work. More formal proofs of the 
validity of operator product expansion order by order in perturbation theory within quantum field theoretic 
models derived from a renormalizable Lagrangian had in fact been established much earlier~\cite{z}.

In general curved spacetimes, a derivation of the operator product expansion from first principles is
not available at present\footnote{Heuristically, one expects that if an operator product expansion 
holds for a theory in Minkowski spacetime, then it also holds for the corresponding theory in curved spacetime, 
since, essentially by the ``Einstein equivalence principle,'' the short distance behavior of a quantum
field theory in curved spacetime should be the ``same'' as that of the corresponding field theory in Minkowski
spacetime.}. In this paper, we will not investigate this important issue, but consider instead the simpler
question which properties of the operator product expansion can be derived in curved spacetime
if one {\em assumes} that such an expansion exists in a suitable sense and that it has certain 
general model-independent properties. Specifically, we are
going to derive the following result about the invariance of the operator product expansion 
under parity, time, and charge (PCT) in a general 4-dimensional analytic curved spacetime:  
If the operator product expansion has the properties\footnote{We emphasize in particular that 
we do not assume that our model is derived from a Lagrangian.} 
(L) that the distributional coefficients $c^{(i)}$ are constructed out of the metric
in a local and covariant manner, (M) that they satisfy a suitable ``microlocal'' spectrum condition~\cite{bfk}, 
(A) that they vary analytically under analytic variations of 
the spacetime metric, then the operator product expansion will automatically have an invariance under 
reversal of the space and time orientations of the spacetime, and charge conjugation of the fields
(if the theory contains any charged fields). Since the coefficients $c^{(i)}$ in the operator
product expansion can be viewed in some sense as ``structure constants'' of the 
algebra of quantum fields, one can interpret this result as showing that the algebraic
structure of the quantum fields in curved spacetime is invariant under PCT at short distances, 
at least under the above general assumptions.
The status of our assumptions is the following: We will argue that property (M) is 
satisfied for the operator in Minkowski spacetime as constructed in~\cite{b}, and we will show elsewhere~\cite{hw3}
that properties (L), (M), and (A) are satisfied for perturbative constructions
of the operator product expansion in an arbitrary curved spacetime.

We would now like to qualify our statement about the PCT invariance of the operator product expansion 
and distinguish it from corresponding statements about (global) PCT invariance. 
For quantum field theories in Minkowski spacetime satisfying the Wightman-axioms
or the axioms of algebraic quantum field theory, one can show~\cite{sw,pj,bo} that PCT is always
implemented by an (anti) unitary operator $\Theta$ on the vacuum Hilbert-space of the theory, in the sense that  
$\Theta \phi(x) \Theta^{-1} = i^F (-1)^M \phi(-x)^*$, where $M$ is the number of unprimed
spinor indices of the field, and where $F$ is zero if the field is bosonic, and one if it is 
fermionic. If the theory has an operator product expansion
like eq.~\eqref{ope}, then one easily finds that this expansion has a similar invariance under PCT
by acting with $\Theta$ on both sides of this expansion. Hence, in Minkowski spacetime, 
the PCT invariance of the operator product expansion is a simple and direct consequence of the 
(global) PCT invariance of the theory. On the other hand,
it is clear that for a quantum field theory defined only on a {\em single, fixed} curved spacetime, 
one cannot in general even formulate the notion of PCT symmetry in the same manner as in Minkowski spacetime, 
since a generic spacetime does not possess any isometries analogous to parity (${\bf x} 
\to -{\bf x}$) and time reversal ($t \to -t$) in Minkowski spacetime. Nevertheless, a notion of (global) 
PCT symmetry can naturally be formulated in a theory that is consistently given on {\em all} 
oriented and time oriented globally hyperbolic spacetimes in the sense of a 
{\it generally covariant quantum field theory} as recently introduced 
in~\cite{bfv,hw1}: 
It is natural to say that a generally covariant quantum field theory is {\em globally PCT invariant} if 
the algebras of observables\footnote{The algebra of observables associated with a given spacetime 
is the abstract *-algebra generated by 
quantum fields smeared with testfunctions of compact support in the spacetime.}
corresponding to a given spacetime equipped with the opposite orientations 
are isomorphic, and if any quantum field 
on the spacetime with the original orientation is mapped to the charge conjugate field on the 
spacetime with the opposite orientation under this isomorphism. 

It is not known at present whether and under what circumstances a generally covariant quantum 
field theory is globally PCT invariant in this sense. Consequently, 
a corresponding PCT invariance of the operator product expansion in curved spacetime does not 
automatically follow in the same straightforward manner as in Minkowski spacetime. As we show in this paper, 
PCT invariance of the operator product expansion can nevertheless be proven 
under the above general and model-independent assumptions (M), (L), and (A). 
As we have already mentioned, this result may be viewed as an ``infinitesimal'' version of the PCT-theorem in 
curved spacetime, in the sense that it proves PCT-invariance of the algebraic 
relations between the quantum fields at short distances.

Our strategy for proving this result is the following. Using that the coefficients in the operator 
product expansion depend locally and covariantly on the spacetime metric 
and the spacetime orientations in a covariant manner, and using that this dependence is analytic, 
we show, using the ideas of~\cite{hw2}, that each coefficient can be expanded into a sum of 
terms, each of which is a product of a curvature tensor at the reference point, $x$, times
a Lorentz-invariant Minkowski space distribution in the Riemannian normal coordinates of the points $y_i$
relative to $x$. The PCT invariance of the coefficients in the operator product expansion 
is then seen to follow if these Minkowski space distributions are invariant (up to permutation
of the arguments and a combinatorical factor) under a reflection of the Riemannian normal coordinates
of the $n$ points $y_i$ about the origin. In order to show that this invariance indeed holds, 
we use the microlocal spectrum condition to show that our Minkowski space distributions arise as 
the boundary value of certain analytic functions. The desired invariance is then shown using the 
transformation properties of these analytic functions under complex Lorentz transformations on 
a suitable complex domain by methods that are similar to the proof of the PCT-theorem in Minkowski spacetime~\cite{sw}. 

We remark that, while our proof makes essential use of the fact that the spacetime is real analytic, 
we expect that our result can be generalized to spacetimes that are only smooth by approximating 
such spacetimes with a sequence of real analytic spacetimes and by making suitable additional 
assumptions about the ``continuity'' of the operator product expansion (of the kind
introduced in~\cite{hw2}) under such approximations. 

The organization of this paper is as follows. In section~2, we recall the notion of a generally covariant quantum field
theory in curved spacetime. In section~3, we give a precise formulation of the operator product expansion in curved
spacetime and in section~4 we state our technical assumptions concerning the properties of this 
expansion.  Section~5 contains the main result of this paper (theorem 5.1). 

Our conventions and notations related to the spacetime geometry are as follows: We view a spacetime as a triple
$\M = (\M, g_{ab}, o)$, where $\M$ is a 4-dimensional manifold, $g_{ab}$ is a metric tensor of signature 
$(+---)$, and $o$ denotes space and time orientations, represented by a tuple $(T, \epsilon_{abcd})$, 
where $T$ is a time function on $\M$ and $\epsilon_{abcd}$ is a nowhere vanishing volume form. Throughout, 
we assume the manifold structure of $\M$ and the spacetime metric $g_{ab}$ to be real analytic. 
We denote (abstract) tensor indices by lower case letters of the Roman alphabet and the components of a 
tensor in a coordinate chart by letters of the Greek alphabet. 

\section{Mathematical formulation of quantum field theory on curved spacetimes}

The usual formulations of quantum field theory on Minkowski spacetime rely heavily on 
the existence of a preferred vaccum state and the special properties of that state. 
The existence of such a state is tied up with the special symmetries of Minkowski 
spacetime, and indeed, there is no preferred state, nor even any preferred Hilbert 
space construction that can be singled out for special consideration on a generic 
curved spacetime. Moreover, apart from a very limited class of spacetimes such as 
static ones, most Lorentzian spacetimes cannot be viewed as a real section of 
a complex spacetime that also possesses a real, Euclidean section, so a formulation 
of quantum field theories on generic Lorentzian spacetimes via Euclidean methods 
such as the Euclidean path-integral is not possible\footnote{By this we do not 
mean that it is not worthwhile to study the Euclidean path integral in curved space, or 
other related quantities, such as e.g. ``effective actions''. What we mean is that the 
physical interpretation of such quantities and their properties is very unclear unless the Euclidean spacetime
under consideration has a real, Lorentzian section. This is a very severe restriction that 
excludes essentially all spacetimes that are not static.} in general. 

Fortunately, there is a simple, and fully satisfactory way to formulate quantum field 
theory in curved spacetime which bypasses all of these problems, namely
the so-called ``(generally covariant) algebraic 
approach to quantum field theory''\cite{bfv, hw1} (for a review of the algebraic 
approach to quantum field theory in Minkowski spacetime, see~\cite{haag}). 
In this framework, a (generally covariant)
quantum field theory is viewed as an assignment that associates with every oriented and time-oriented 
spacetime $\M \equiv (\M, g_{ab}, o)$ 
an abstract *-algebra\footnote{In~\cite{bfv}, these algebras were assumed to 
be C$^*$-algebras. This is to restrictive for the purposes of the present paper since
we also want $\cA(\M)$ to contain unbounded elements.} $\cA(\M)$ with unit whose elements are the observables of 
the theory. The features of locality and general covariance of a quantum field theory are 
reflected in the following consistency properties of this assignment: 

Consider a situation in which we are given an isometric embedding 
$\chi: \N \to \M$ of a spacetime $\N$ into 
a spacetime $\M$ which preserves the causal structure and the orientations, 
meaning that if $o = (T, \epsilon_{abcd})$ is the orientation of $\M$, then 
$(\chi^* T, \chi^* \epsilon_{abcd})$ coincides with the orientation of $\N$. 
Then we postulate that there exists an injective *-homomorphism
\begin{equation}
\label{alphadef}
\alpha_\chi: \cA(\N) \to \cA(\M). 
\end{equation}
Furthermore, if 
$\chi_1$ and $\chi_2$ are such isometric embeddings with the above properties
between spacetimes such that the composition $\chi_1 \circ \chi_2$ can be defined (and consequently 
defines again an isometric embedding with the these properities), then 
we have 
\begin{equation}\label{comp}
\alpha_{\chi_1 \circ \chi_2} = \alpha_{\chi_1} \circ \alpha_{\chi_2}.
\end{equation}
The existence of the algebraic isomorphism $\alpha_\chi$ in eq.~\eqref{alphadef}
with the property~\eqref{comp} formalizes the idea that observables associated
with a spacetime $\N$ that is isometric to a globally hyperbolic subregion
of a larger spacetime $\M$ can be viewed via $\alpha_\chi$ as observables in the 
larger spacetime satisfying the same algebraic relations, which can be interpreted as 
saying that the
algebraic relations between the observables depend locally and covariantly on the metric.
If $\M$ is Minkowski spacetime with a given choice of orientations, 
then the orientation and causality preserving (global) isometries 
of $\M$ are given precisely by the translations 
$x \to x + a$, where $a \in \mr^4$, together with the proper orthochronous Lorentz 
transformations $x \to \Lambda x$, where $\Lambda \in \cL$. Thus, in the special 
case of Minkowski spacetime, our axioms say that the Poincare group acts
on the algebra of observables by a group of *-automorphisms 
$\alpha_{\{\Lambda, a\}}$. Requirements eq.~\eqref{alphadef}
and eq.~\eqref{comp} may therefore be viewed as a replacement of the notion of
Lorentz-covariance of a quantum field theory 
in Minkowski spacetime by the notion of general covariance. 

In order to formulate the notion of local commutativity respectively local anticommutativity in this
algebraic framework, we need to assume that algebraic elements $A \in \cA(\M)$ can be uniquely decomposed 
into a ``bosonic'' and a ``fermionic'' part. This is formalized by requiring that  
there exists a *-automorphism $\gamma_\M$ for every oriented spacetime $\M$ with 
the property $(\gamma_\M)^2 = 1$ and $\gamma_\M = \gamma_\N \circ \alpha_\chi$
whenever $\chi$ is an orientation and causality preserving isometric embedding from $\N$ into
$\M$. We can then uniquely decompose $A = A_+ + A_-$, where $\gamma_\M(A_\pm) = \pm A_\pm$, and 
we call $A_+$ the bosonic and $A_-$ the fermionic part of $A$. 
Given now two isometric embeddings $\chi_i: \N_i \to \M$, $i = 1, 2$, such that 
the image of $\N_1$ under $\chi_1$ in $\M$ is spacelike related to the image 
of $\N_2$ under $\chi_2$, then our requirement of local (anti-) commutativity is
\begin{equation}
\label{com}
[\alpha_{\chi_1}(A_1), \alpha_{\chi_2}(A_2)]_\gamma = 0
\end{equation} 
for all $A_1 \in \cA(\N_1)$ and $A_2 \in \cA(\N_2)$, where 
\begin{equation}
[A, B]_\gamma = 
\begin{cases}
AB + BA & \text{$A, B$ fermionic,}\\
AB - BA & \text{$A$ or $B$ bosonic,}
\end{cases}
\end{equation}
is the graded commutator.  

The algebras of observables, $\cA(\M)$, were referred to as ``abstract'', 
because it has not been assumed that its elements are represented as linear operators 
on some particular Hilbert space. This is of great conceptual advantage, because there exist in general 
many inequivalent representations of which no particular one can be singled out for 
special consideration. The quantum states are simply all 
linear functionals $\omega: \cA(\M) \to \mc, A \to \langle A \rangle_\omega$ from the algebra of 
observables associated with that spacetime into the complex numbers, 
which are positive in the sense that $\langle A^*A \rangle_\omega \ge 0$ for all $A \in \cA(\M)$, 
and which are normalized in the sense that $\langle \myid \rangle_\omega = 1$, where $\myid$ is
the identity element. By formulating the theory in terms of abstract algebras, 
we have therefore avoided predjudicing ourselves towards the particular class of 
states that can be represented as vectors or density matrices in some particular 
representation. States of particular interest may be singled out for example
in spacetimes which happen to have symmetries or suitable asymtotic regions or
in models with additional internal symmetries, but we emphasize that 
the question whether such choices are possible is not 
in any way related to the formulation of the quantum field theory. 

A {\em local covariant (scalar) field}, $\phi$, is an assignment which associates with every spacetime
$\M$ a linear map 
\begin{equation}
\phi_\M: \cD(\M) \to \cA(\M), \quad f \to \phi_\M(f), 
\end{equation}
from the space $\cD(\M)$ of all smooth compactly supported functions on $\M$ to $\cA(\M)$, 
which satisfies 
\begin{equation}\label{lcf}
\alpha_\chi(\phi_\N(f)) = \phi_\M(\chi_* f), 
\end{equation}
whenever $\chi: \N \to \M$ is an orientation and causality preserving isometric 
embedding of a spacetime $\N$ into a spacetime $\M$, and where $\chi_* f$ denotes
the testfunction on $\M$ corresponding to the testfunction $f$ on $\N$ 
via the map $\chi$. The above transformation law~\eqref{lcf}
expresses (a) that the field $\phi$ is constructed entirely out of the metric in an 
arbitrary small neighborhood of the point $x$, and (b) that it is constructed 
out of the metric in a generally covariant way. In the case when $\M$ is Minkowski 
spacetime and $\chi = \{\Lambda, a\}$ is an element of the Poincare group, 
eq.~\eqref{lcf} specializes to $\alpha_{\{\Lambda, a\}} (\phi(x)) = \phi(\Lambda x + a)$, 
which is the familiar special relativistic transformation law for a scalar field.  

The above definition of local covariant quantum field of scalar type can be generalized
in a relatively straightforward manner to fields of arbitrary spinor type. The main new
issue is that the definition of spinors curved spacetime requires the existence and specification of 
a spin structure. We will consequently assume that the spacetimes under consideration 
can be equipped with a spin structure (for matters related to spinors in curved 
spacetime, see appendix B). 
Since we want the quantum fields of spinor type to be elements in $\cA(\M)$ after smearing 
with a suitable testfunction, we will now view these algebras as depending not only on 
the spacetime metric and orientations, but also on the particular 
choice of spin structure, if several inequivalent spin structures are possible. 
The above locality and covariance property~\eqref{alphadef} and the local (anti-) commutativity
property~\eqref{com} of the assignment $\M \to \cA(\M)$
are then formulated in terms of embedding maps $\chi: \N \to \M$ which not only  
preserve the metric structure and orientation, but in addition also lift to 
a homomorphism between the spin structures on $\N$ respectively $\M$. With these modifications
understood, local and covariant quantum fields of spinor type are then defined
in precisely the same way as in the scalar case, with the only difference 
that the space of test functions $\cD(\M)$ is now replaced by the appropriate space of 
smooth test spinors whose elements are compactly supported 
smooth sections in vector bundles $\cV(\M)$, $\cV'(\M)$, $\cV^{\prime *}(\M)$ and $\cV^*(\M)$
(corresponding to unprimed, primed, upper primed, and upper unprimed spinor indices)
that are associated with the spin structure over $\M$. In other words, 
a local covariant quantum field of spinor type is an assignment
\begin{equation}
\phi_\M: \cD(\M; \cF(\M)) \to \cA(\M), \quad f \to \phi_\M(f), 
\end{equation}
valued in the algebra of observables $\cA(\M)$, where $\cF(\M)$ is 
a suitable tensor product of the bundles 
$\cV(\M)$, $\cV'(\M)$, $\cV^{\prime *}(\M)$ and $\cV^*(\M)$ corresponding 
to the spinor type of the field $\phi$. If $\chi$ is an isometric embedding from another 
oriented, time oriented spacetime $\N$, that 
preserves causality, orientation and time orientation and which lifts to a 
corresponding map between the respective spinor structures, then  
\begin{equation}
\alpha_\chi(\phi_\N(f)) = \phi_\M(\chi_* f)
\end{equation}
holds, where $\alpha_\chi: \cA(\N) \to \cA(\M)$ is as in eq.~\eqref{alphadef}, and where
$\chi_* f$ denotes the testfunction in $\cD(\M, \cF(\M))$ which is naturally associated
with $f$ via the identification of the spin structures over $\M$ and $\N$ given by $\chi$.

For the quantum field theories that we consider in this paper, we assume that there are 
countably many local covariant fields, which we shall denote by the generic symbol
$\phi^{(i)}$, where $(i) \in {\mathbb N}$ is a label that distinguishes the various fields.

\medskip

We note that, since the grading maps $\gamma_\M$ satisfy $\gamma_\M = \gamma_\N \circ
\alpha_\chi$ for every orientation and causality preserving isometric embedding, we can 
consistently decompose any local and covariant field into its bosonic and fermionic 
parts for all spacetimes. Thus, without loss of generality, we can assume that a local covariant field is either
bosonic or fermionic. We emphasize however that we do not assume that that all
half odd-integer spin fields are fermionic and that all integer spin fields are bosonic.
Such a relation between spin and statistics has been proven recently by Verch~\cite{ver}, 
but the technical assumptions made in \cite{ver} are not identical with the technical
assumptions we will be making here. The proof of our main result on the other hand does not 
rely on the spin-statistics relation, so we will avoid assuming the spin-statistics 
relation in this paper. 

\section{Formulation of the operator product expansion in curved spacetime}

In the last section we have reviewed the formulation of quantum field theory in 
curved spacetime as an assignement of spacetimes with *-algebras of observables,
and we have introduced local, covariant quantum fields as suitable assignments 
of spacetimes with elements in the algebra of observables associated with the spacetime. 
We now wish to study quantum field theories in curved spacetime
that possess in addition an operator product expansion.

Let $\Sigma(\M)$ of all complex linear functionals on $\cA(\M)$, 
\begin{equation}
\Sigma(\M) = \{\sigma: \cA(\M) \to \mc \mid \sigma(c_1 A_1 + c_2 A_2) 
= c_1 \sigma(A_1) + c_2 \sigma(A_2) \}.  
\end{equation}
We say that such a functional is {\em real}, if $\sigma(A^*) = \overline{\sigma(A)}$ for all 
$A$. Quantum states $\omega: A \to \langle A \rangle_\omega$ 
are normalized and positive elements of $\Sigma(\M)$.

The proof~\cite{b} of the operator product expansion in Minkowski
spacetime suggests that one should view the coefficients $c^{(i)}$ appearing in the operator
product expansion~\eqref{ope} as being the $n$-point functions of certain ``standard'' 
linear functionals $\sigma^{(i)}$ on $\cA(\M)$, where $(i)$ is a label that distinguishes 
the various local and covariant fields in the theory. We will adapt this viewpoint in our formulation 
of the operator product expansion in curved spacetime. 
Our (as yet, still formal) definition of a local, covariant quantum field theory possessing 
an operator product expansion is then as follows:

\begin{defn}
\label{opedef}
We say that a local covariant quantum field theory (with only scalar fields) possesses
an operator product expansion, if for any space and time oriented spacetime 
$\M$ and any point $x \in \M$ there exist linear functionals 
\begin{equation}
\label{siggi}
\sigma^{(i)}_{\M, x} \in \Sigma(\M)  
\end{equation}
such that
\begin{equation}
\label{f}
\sigma_{\M, x}^{(i)} \circ \gamma_\M = (-1)^{F^{(i)}} \sigma_{\M, x}^{(i)}, 
\quad F^{(i)} =
\begin{cases}
0 & \text{if $\phi^{(i)}$ is bosonic,}\\
1 & \text{if $\phi^{(i)}$ is fermionic,}
\end{cases}
\end{equation}
and
\begin{equation}
\label{ope1}
\Big\langle \prod_{k=1}^n \phi^{(j_k)}_\M(y_k) \Big\rangle_\omega - 
\sum_{i = 1}^N \sigma^{(i)}_{\M, x}\Big(\prod_{k=1}^n \phi^{(j_k)}_\M(y_k)\Big)\, \Big\langle \phi^{(i)}_\M(x) \Big\rangle_\omega
\to 0, 
\end{equation}
as $(y_1, \dots, y_n) \to (x, \dots, x)$ and as $N \to \infty$, for all suitable states $\omega$ 
on $\cA(\M)$, and any collection of fields $\phi^{(j_1)}, \dots, \phi^{(j_n)}$. 
\end{defn}
\paragraph{\bf Remarks:}
(1) The coefficients $c^{(i)}$ in our previous expression for the operator product expansion~\eqref{ope} correspond
to the standard functionals $\sigma^{(i)}$ in the above formulation via
\begin{equation}
\label{cidef}
c^{(i)}_{\M, x}(y_1, \dots, y_n) = \sigma^{(i)}_{\M, x}(\phi^{(j_1)}_\M(y_1) \cdots \phi^{(j_n)}_\M(y_n)), 
\end{equation}
where we have now put a subscript ``$\M, x$'' on the coefficients $c^{(i)}$ in order to indicate the 
dependence on the spacetime and the reference point, $x$. Condition eq.~\eqref{f} expresses the 
demand that each term in the operator product expansion has the same fermion number modulo 2. 

\medskip
\noindent
(2) The above definition can be generalized in a straightforward way to theories that contain
not only scalar fields but fields of arbitrary spinor type, in which case all quantities depend 
in addition on a choice of spin structure over $\M$ which is compatible with the space and 
time orientations (see appendix B for details). Since local covariant fields  
of spinor type take testfunctions that are sections in a vector bundle $\cF(\M)$ corresponding to the spinor 
type of the field, it is natural in this case to view $\sigma^{(i)}_{\M, x}$ as 
linear functionals on $\cA(\M)$ taking values not in $\mc$ but instead in the complex vector space $\cF_x(\M)$, 
where we mean the fibre of this vector bundle over $x$.

\medskip
\noindent
To make the above definition mathematically precise, we still need to specify
\begin{enumerate}
\item[(a)] 
the precise nature of the states $\omega$ that are allowed in eq.~\eqref{ope1}, as well as 
the nature of the functionals $\sigma_{\M, x}^{(i)}$.  
\item[(b)] the precise sense in which 
the expression~\eqref{ope1} tends to 0. 
\end{enumerate}
We now turn to these tasks.  

Given a spacetime $\M$, a collection $\phi^{(j_1)}, \dots, \phi^{(j_n)}$ of local 
covariant fields and a functional $\sigma \in \Sigma(\M)$, we consider the multi-linear functional
\begin{equation}
\label{npt}
\times^n \cD(\M) \to \mc, \quad (f_1, \dots, f_n) \to \sigma(\phi^{(j_1)}_\M(f_1) \cdots \phi^{(j_n)}_\M(f_n))
\end{equation}
on the $n$ fold cartesian product of the space of testfunctions on $\M$, where for simplicity we 
assume that all the fields are scalar. The regularity properties of a 
functional $\sigma$
may be specified by specifying regularity properties for the linear functionals~\eqref{npt}
for an arbitrary set of local covariant fields. Firstly, we will ask that the linear
functionals~\eqref{npt} are distributions on $\times^n \M$, i.e., that they are continuous with respect to 
the Laurent-Schwarz topology on the spaces of testfunctions\footnote{Strictly speaking, we should 
demand that our functionals are continuously defined on the space $\cD(\times^n \M)$ rather than 
continuous multilinear functionals on $\times^n \cD(\M)$. However, by the ``Schwartz Nuclear Theorem'', 
these requirements are actually equivalent.} $\times^n \cD(\M)$. Among these, 
we now further restrict our attention to those functionals $\sigma$ for which the distributions~\eqref{npt}
have a particular singularity structure specified by the following ``microlocal spectrum condition''~\cite{bfk}:
\begin{equation}
\label{wfc}
\WF_A(\sigma(\phi^{(j_1)}_\M(y_1) \cdots \phi^{(j_n)}_\M(y_n))) \subset \Gamma_\M, 
\end{equation} 
where $\WF_A$ is the ``analytic wave front set''~\cite{h} of a distribution\footnote{
Our convention for the Fourier transform in $\mr^m$ is $\hat f(k) = (2\pi)^{-m/2} \int \e^{+ikx} f(x) \, d^m x$, which
is opposite to the convention used in~\cite{h}. It follows from this that our definition of the analytic 
wave front set is minus the definition given in~\cite{h}.}, 
and where $\Gamma_\M \subset T^*(\times^n \M) \setminus \{0\}$ is defined in terms of the geometry as follows: 
Let $G(p)$ be a ``decorated embedded graph''
in $\M$. By this we mean an embedded graph in $\M$ whose 
vertices are points $x_1, \dots, x_n$ in $\M$ 
and whose edges, $e$, are piecewise smooth curves\footnote{
We note that a more restrictive notion of a microlocal spectrum condition would be 
obtained if we would replace ``piecewise smooth curve'' by ``causal curve'' or 
``null-geodesic''.}
$\gamma$ in $\M$ connecting the vertices. Each such 
edge $e$ is equipped with a future pointing timelike or null coparallel covectorfield $(p_e)_a$, meaning 
that 
\begin{equation}
\dot \gamma^a \nabla_a (p_e)_b = 0, \quad g^{ab} (p_e)_a (p_e)_b \ge 0, \quad (p_e)^a \nabla_a T > 0, 
\end{equation}
where $T$ is the time function that defines the time orientation of $\M$.   
If $e$ is an edge in $G(p)$ connecting the points $x_i$ and $x_j$ 
with $i < j$, then we denote $s(e) = i$ its source 
and $t(e) = j$ its target. With this notation, we define
\begin{eqnarray}
\label{gamtdef}
\Gamma_\M &=& 
\Big\{(x_1, k_1; \dots; x_n, k_n) \in T^*(\times^n \M)  \setminus \{0\} \mid 
\exists \,\, \text{decorated graph $G(p)$ with vertices} \nonumber\\
&& \text{$x_1, \dots, x_n$ such that
$k_i = \sum_{e: s(e) = i} p_e - \sum_{e: t(e) = i} p_e 
\quad \forall i$} \Big\}. 
\end{eqnarray}
We will denote by 
\begin{equation}
\Sigma_A(\M) = \{ \sigma \in \Sigma(\M) \mid \WF_A(\sigma(\prod_{k = 1}^n \phi_\M^{(j_k)}(y_k))) 
\subset \Gamma_\M \}
\end{equation}
the space of all linear functionals such that eq.~\eqref{wfc} holds for an arbitrary set of local covariant fields. 
Our operator product expansion will be required to hold only for states $\omega \in \Sigma_A(\M)$. 
The analytic wave front set of $\langle \phi^{(i)}(x) \rangle_\omega$ is then empty, 
meaning that this expression is not just a distribution, but in fact an analytic function in $x$. 
This implies in particular that the products of distributions implicit in our operator product expansion~\eqref{ope1}
are automatically well-defined. 

We furthermore require that the 
standard functionals $\sigma^{(i)}_{\M, x}$ are such that eq.~\eqref{wfc} is satisfied in some neighborhood
of $x$; in other words, we require:
\begin{enumerate}
\item[(M)]
There exists an isometric embedding $\chi: \N \to \M$ preserving the
orientations such that the linear functional on $\cA(\N)$ defined by 
\begin{equation}
\cA(\N) \owns A \to \sigma_{\M, x}^{(i)}(\alpha_\chi(A)) \in \mc
\end{equation}
is an element of $\Sigma_A(\N)$. 
\end{enumerate}
We have thus accomplished (a). 

The above microlocal spectrum condition (or rather, an analogous ``$C^\infty$''-version thereof) 
was first proposed by \cite{bfk}, as a replacement for the usual spectrum condition on vacuum
states in Minkowski spacetime. It was shown in~\cite{bfk} that it is satisfied 
in any Wightman quantum field theory in Minkowski spacetime, as well as for 
so-called quasifree ``Hadamard states''~\cite{kw} in linear quantum field theories in curved 
spacetimes\footnote{In fact, the Hadamard form of the two-point function can be shown to be
{\em equivalent} to a suitably strengthened version (see footnote 9 on p.~11) of the $C^\infty$-microlocal spectrum 
condition~\cite{rad} (for a discussion of the analytic case, see~\cite{ver1}).}. 
The above analytic version of this condition is natural in 
analytic spacetimes and was first proposed in~\cite{hw1}. It is discussed in~\cite{ver1} in 
connection with long-range correlations in quantum field theories.

Our motivation for imposing the microlocal spectrum condition~\eqref{wfc} on the operator 
product expansion comes from the following facts. It was shown in \cite{b} that the operator product expansion 
in Minkowski spacetime will hold typically only for states $\omega$ that are well-behaved at high energies (for example 
energy-bounded), and that the standard functionals $\sigma_{\M, x}^{(i)}$ can be chosen energy-bounded. 
On the other hand, one can show that if $\M$ is Minkowski spacetime, then every functional 
with bounded energy satisfies the microlocal spectrum condition. 
More specifically, assume that the algebra of observables corresponding to Minkowski space 
admits a faithtfull representation on a Hilbert space on which the 
group of automorphisms $\alpha_{a}$ associated with the translations by a four vector $a$ 
is implemented by a strongly continuous group of unitaries, $\alpha_{a}(A) = 
\e^{ia^\mu P_\mu} A \e^{-ia^\mu P_\mu}$, with self-adjoint generator $P$ satisfying the 
spectrum condition, ${\rm spec} P \subset \bar V^+$, where $\bar V^+$ is the closure of 
the future lightcone in Minkowksi spacetime, and where 
$A$ has been identified with the linear operator on the Hilbert space representing it. 
We say that a functional $\sigma \in \Sigma(\M)$ in Minkowski space has finite energy below $p^0$ 
(relative to some Lorentz frame) if 
\begin{equation}
\label{bde}
\sigma(A) = \sigma(E_{p^0} A E_{p^0}) \quad \forall A \in \cA(\M), 
\end{equation}
where $E_{p^0}$ denotes the projector on the spectral subspace of the Hamiltonian $P^0$ corresponding
to energies less than $p^0$, and where we have assumed that $\sigma$ can be identified with a functional
on the image of $\cA(\M)$ under the represenation. Then such a $\sigma$
satisfies the microlocal spectrum condition~\eqref{wfc} in Minkowski spacetime
(a formal proof of this statement, which follows closely a similar argument invented in~\cite{bfk}, is given in 
appendix A). Furthermore, it seems to be the case that the coefficients
in the operator product expansion for free fields in analytic curved spacetimes satisfy our analytic 
microlocal spectrum condition~\cite{hw3}, and we expect this also to be true for perturbatively 
defined self-interacting quantum field theories in curved spacetimes.
  
\medskip

We next turn to our second task (b) to explain the precise sense in which 
the expression~\eqref{ope1} converges to 0. An investigation of the operator product 
expansion for free fields shows that one can certainly not expect that expression~\eqref{ope1} tends to zero in 
the sense of a convergent sequence of functions, or rather, distributions. Rather, one can 
only expect that this expression has an arbitrarily low scaling degree as $(y_1, \dots, y_n)
\to (x, \dots, x)$ for large $N$ in any given state $\omega \in \Sigma_A(\M)$. The model-independent
derivation~\cite{b} of the operator product expansion in Minkowski spacetime 
from first principles leads to the same conclusion\footnote{We remark however that the 
convergence properties of the operator product expansion established in~\cite{b} are 
stronger than the convergence properties postulated here in that they hold {\em uniformly}
for all states with energy below some arbitrary $p^0$.}. 
We will consequently formulate the convergence of the 
operator product expansion by demanding that the scaling degree of the difference~\eqref{ope1}
becomes arbitrarily small when $N \to \infty$. 

Let $u$ be a distribution on an open, convex neighborhood $X$ of $\mr^n$. If $\lambda$
is a positive number less than 1 and $f$ is a smooth compactly supported function on $X$, 
we define another such function $f_\lambda$ by setting $f_\lambda(y) = \lambda^{-n} f(x + \lambda(y-x))$. 
The scaling degree~\cite{st}, $\delta$, of $u$ at the point $x$ is defined as 
\begin{equation}
\delta = \inf \{ \gamma \in \mr^+ \mid \lim_{\lambda \to 0} \lambda^\gamma u(f_\lambda) = 0 \quad \forall
f \in \cD(X) \}.
\end{equation}
The scaling degree of a distribution thus characterizes the strength of its singularity at $x$.  
It is a completly local concept in that it depends only on the behaviour of $u$ near $x$ and 
can be generalized in an invariant manner to distributions on a manifold $X$ by 
localizing $u$ in a chart near a point $x$ in the manifold.  

The precise sense in which we assume the operator product expansion to converge is then the following:
We ask that for every $\delta < 0$, we can find an $N$ such that the
scaling degree of the distribution defined by the left side of expression~\eqref{ope1} at $(y_1, \dots, y_n)
= (x, \dots, x)$ is less than $\delta$. This accomplishes (b). We note however that the PCT-invariance of 
the operator product expansion that we are going to state and prove in section~5 will follow independently
of any assumptions made about the convergence of this expansion at small distances---in other words, property 
(b) will not be used at all in the proof given in section~5 (see also the remark following theorem~\ref{mainthm}).

\section{Technical Assumptions about the OPE}

In the last section we have given a mathematically precise formulation of the operator 
product expansion in a curved spacetime. In order to be able to prove our main result that 
the operator product expansion~\eqref{ope1}
has a PCT-invariance, we will now make the following further assumptions about the nature of this 
expansion:
\begin{enumerate}
\item[(L)] The standard functionals $\sigma^{(i)}_{\M, x}$ have a local and covariant dependence on the 
spacetime metric and orientations.  
\item[(A)] The standard functionals $\sigma^{(i)}_{\M, x}$ have a suitable analytic
variation under analytic variations of the spacetime metric.
\end{enumerate}  
Our motivation for imposing (L) and (A) comes from the fact that these 
properties are satisfied in free field theories in curved spacetime and 
are also expected to hold in perturbatively defined interacting quantum 
field theories in curved spacetime~\cite{hw3}. As we explain below, condition (L) is 
equivalent to the local and covariant dependence on the metric of the coefficients $c^{(i)}$
in the operator product expansion~\eqref{ope}. Since these coefficients can be viewed, 
in some sense, as structure constants for the algebraic relations between the quantum 
fields at short distances, we may view (L) as a strengthened version of the general
covariance property of the quantum field theory under consideration. 
We now discuss the precise form (L) and (A) in turn. 

In order to formulate our condition that the standard functionals depend locally and covariantly 
on the metric, it is useful to first define an equivalence relation $\xsim$ between 
linear functionals in $\Sigma(\M)$ relative to a point $x \in \M$
by declaring two such functionals $\varphi_1$ and $\varphi_2$ to be equivalent if they coincide when restricted to 
some neighborhood of the point $x$, where the restriction of a linear functional on $\cA(\M)$
to a globally hyperbolic neighborhood $\cO \subset \M$ is defined in the obvious way 
by viewing $\cA(\cO)$ as a subalgebra of $\cA(\M)$ via the *-isomorphism eq.~\eqref{alphadef}
corresponding to the embedding $\cO \subset \M$. The assignment of oriented, time oriented 
spacetimes $\M$ with functionals $\sigma^{(i)}_{\M, x}$ is then said to be local and covariant if 
\begin{equation}
\label{lcv'}
\sigma^{(i)}_{\M, \chi(x)} \circ \alpha_\chi \xsim \sigma^{(i)}_{\N, x}
\end{equation}
for any orientation and causality preserving isometric embedding $\chi: \N \to \M$, where we 
have assumed for simplicity that all fields in the theory are scalar. 
In the case when the theory contains spinor fields as well, we consider causality and 
orientation preserving isometric embeddings $\chi$ that in addition lift to a corresponding
map between the spin-structures over $\N$ and $\M$ respectively. If $\cF(\N)$ is the vector bundle
over $\N$ corresponding to the spinor type of the field $\phi^{(i)}$, then, as explained in the remark
following def.~\ref{opedef}, the functional $\sigma^{(i)}_{\N, x}$ should be viewed as taking values 
not in $\mc$, but in the finite dimensional vector space $\cF_x(\N)$, and likewise for the 
functional $\sigma^{(i)}_{\M, \chi(x)}$. 
The analog of eq.~\eqref{lcv'} for spinor fields is then
\begin{equation}
\label{lcv-}
\sigma^{(i)}_{\M, \chi(x)} \circ \alpha_\chi \xsim \chi_* \sigma^{(i)}_{\N, x},  
\end{equation}
where $\chi_*: \cF_x(\N) \to \cF_{\chi(x)}(\M)$ is the linear map induced by $\chi$. 
The above locality and covariance conditions~\eqref{lcv'} and~\eqref{lcv-} imply that the $n$-point functions~\eqref{cidef} of 
our standard functionals are distributions which are locally and covariantly constructed out of the metric and
the orientations near the reference point, $x$. Namely if $\chi: \N \to \M$ is an 
orientation and causality preserving isometric embedding, then it follows immediately from 
the transformation law of the fields~\eqref{lcf} and the functionals~\eqref{lcv'} that 
\begin{equation}
\label{lcv''}
\chi^* c^{(i)}_{\M, \chi(x)}(y_1, \dots, y_n) = c^{(i)}_{\N, x}(y_1, \dots, y_n)
\end{equation}
in the sense of distributions for all $y_j$ in some neighborhood of the point $x$, where 
$\chi^*$ denotes the pull-back of a distribution, defined by analogy with the pull back 
of a smooth density. The reader may wonder why we are not demanding equality in eq.~\eqref{lcv'}
rather than only equivalence under $\xsim$, or alternatively, why we do not impose that 
relation~\eqref{lcv''} holds for all $y_j$ in $\N$, rather than some neighborhood of the point
$x$. The reason for this is that we typically expect the coefficients~\eqref{cidef} to 
contain expressions like the geodesic distance, $s_\M(y_1, y_2)$, between two points in $\M$ 
near $x$. Now the geodesic distance between two points is {\em not} a quantity that is 
locally constructed out of the metric, since the geodesic distance between two points in a 
spacetime $\N$ (even if it can be defined unambiguously) can be made shorter by embedding 
$\N$ into a suitably chosen larger spacetime $\M$. Therefore, it is not true that 
$\chi^* s_\M = s_\N$ for the geodesic distance. On the other hand, it is true that $\chi^* s_\M
= s_\N$ when both sides are restricted to suitably small neighborhood $\cO$ of $x$.  

Our locality and covariance condition as stated above requires only that there is 
{\it some} region, $\cO$, such that both sides of~\eqref{lcv'} are equal upon restriction
to $\cO$, but we have not imposed any requirements upon the size of $\cO$, which could vary 
arbitrarily so far as the embedding varies. For technical reasons, we must also impose 
the additional condition that $\cO$ can be chosen uniformly in the following sense as the embedding $\chi$ varies. 
We ask that for every spacetime $\M$ and point $x$, 
there exists an open neighborhood $\mathcal X$ of the identity in the space ${\rm Diff}_x^A(\M)$
of analytic diffeomorphisms on $\M$ leaving $x$ fixed such that $\sigma^{(i)}_{\M, x} \circ \alpha_\chi
= \sigma^{(i)}_{\chi^* \M, x}$ for all $\chi \in \mathcal X$, when restriced to some {\em fixed} 
$\cO$. We view this additional requirement as part of our definition of locality and covariance
of the functionals in the operator product expansion. 

\medskip

We next want to formulate condition (A) that the local, covariant  
functionals in the operator product expansion have an analytic dependence 
under analytic variations of the spactime metric. For this, we consider 1-parameter
families of real analytic metrics $g_{ab}^{(s)}$ on $\M$ which vary analytically with 
respect to a real parameter $s \in I = (a, b)$ in the sense that 
\begin{equation}
g_{ab}^{(s)}-(ds)_a^{}(ds)_b^{} 
\end{equation}
is a real analytic metric on the real analytic 5-dimensional manifold
$I \times \M$. Since the standard functionals in the operator product expansion 
have already been assumed to be locally and covariantly constructed out of the 
metric, we therefore obtain from the family of metrics $g_{ab}^{(s)}$ 
a corresponding family of functionals (labelled by the parameter $s$) 
associated with this family of metrics. 
Our analyticity requirement (A) is then, in essence, that all $n$-point functions
of these functionals have a suitable analytic dependence on the parameter $s$. 

A complication arises from the fact that these $n$-point functions 
are not analytic functions but rather only distributions, so we must first 
consider the question what we actually mean by the statement that a family of distributions 
depends analytically on a parameter. Following~\cite{hw2}, we make the following 
definition.
\begin{defn}
We say that a family of distributions $u^{(s)}$ on 
an analytic manifold $X$ depends analytically on the parameter $s \in I = (a, b)$ with 
respect to a family of conic sets $K^{(s)} \subset T^*(X) \setminus \{0\}$ 
if (a) the dependence on $s$ of the family of distributions $u^{(s)}$ on $X$ 
is such that can be viewed as a distribution $\tilde u$  
on $\tilde X = I \times X$ and if (b) it holds that
\begin{equation}
\label{wfau}
\WF_A(\tilde u) \subset \{(\tilde x, \tilde k) \in T^*(\tilde X) 
\setminus \{0\} \mid f^{(s)}(x) = \tilde x, \quad (x, {^t f^{(s)\prime}(x)} \tilde k) \in K^{(s)} \},  
\end{equation}
where $f^{(s)}: X \to \tilde X$ maps any point $x \in X$ to $\tilde x = (s, x) \in \tilde X$, 
$f^{(s) \prime}$ is the differential of this map viewed as a linear map $T(X) \to T(\tilde X)$, 
and ${^t f^{(s)\prime }}$ denotes the transpose of this linear map, acting between $T^*(\tilde X) \to T^*(X)$. 
\end{defn}
A detailed discussion and motivation of this definition is given in 
\cite[app. A]{hw2}; here we only note the following facts. Firstly, if $\tilde u \in \cD'(\tilde X)$
is any distribution satisfying~\eqref{wfau}, then by the results of~\cite[thm. 8.5.1]{h}, 
the pull-back of this distribution and all of its $s$-derivatives 
by the map $f^{(s)}$ exists as a distribution on $X$ for any $s \in I$ 
and defines an analytic family $u^{(s)}$ of distributions in the sense of the above definition, with each member
satisfying $\WF_A(u^{(s)}) \subset K^{(s)}$. In the special case when the cones $K^{(s)}$ are empty for all $s$, 
we consequently have that $\WF_A(u^{(s)}) = \emptyset$, so each $u^{(s)}$ 
is an analytic function on $X$. The set~\eqref{wfau} is then empty as well 
and the family is consquently jointly analytic in $s$ and $x$. Thus, when $K^{(s)}$ is empty, 
our definition of the analytic dependence on a parameter coincides with the natural 
notion for analytic functions. 

With this definition in mind, we now state the precise form of condition (A).
Let $(\M, g_{ab}^{(s)})$ be 
a family of analytic spacetimes whose metrics vary analytically with $s$, and suppose 
that there is a corresponding analytic family of time functions $T^{(s)}$ and volume forms 
$\epsilon_{abcd}^{(s)}$ for all $s$, which thus define a family of space and time oriented 
spacetimes $\M(s)$.
We say that $\sigma^{(i)}_{\M, x}$ depend analytically on the 
metric if there is a neighborhood $\N$ of $x$ such that the restriction of 
\begin{equation}
\label{sfunc}
(f_1, \dots, f_n) \to \sigma^{(i)}_{\M(s), x}(\phi^{(j_1)}_{\M(s)}(f_1)
\cdots \phi^{(j_n)}_{\M(s)}(f_n))
\end{equation}
to $\times^n \cD(\N)$ is a family of distributions that depends analytcally
on $s$ with respect to the family of conic sets $\Gamma_{\M(s)}$
defined in eq.~\eqref{gamtdef} for every set of local, covariant 
fields $\phi^{(j_1)}, \dots, \phi^{(j_n)}$.

\section{PCT-invariance of the operator product expansion}

We are now going to formulate our main result about the PCT invariance
of the operator product expansion in curved spacetime. Let $\M$ 
be a globally hyperbolic spacetime with metric $g_{ab}$ and space-time
orientation $o = (T, \epsilon_{abcd})$, which admits 
a spin-structure. Let $\overline \M$ be the spacetime whose manifold structure and 
metric coincides with that of our original spacetime, but whose space and time 
orientation is given by $-o = (-T, \epsilon_{abcd})$, i.e., 
are reversed relative to those of the original spacetime. 

Since the definition of spinors involves a choice of orientation, the notion of 
spinors on $\M$ and $\overline \M$ will not coincide. Therefore, in order to formulate
a relation between the operator product expansions on $\M$ and $\overline \M$ involving
spinors, one needs to identify spinors on $\M$ with spinors on $\overline \M$. As we show in 
the appendix B, it is always possible to choose the spinor structures on $\M$ and
$\overline \M$ in such a way that a natural identification is possible, namely, we get a map
\begin{equation}
\tilde I: \cV(\M) \to \cV(\overline \M)
\end{equation}
between the corresponding associated vector bundles of which the spinors are elements. 
In the following, we shall therefore always assume that the spin structures over $\M$ 
and $\overline \M$ have been chosen so that such an identification is possible. 
The same remarks apply to the bundles $\cV^*(\M), \cV'(\M), \cV^{\prime *}(\M)$, as well as their tensor products. 

\begin{thm}
\label{mainthm}
Suppose that a local, covariant quantum field theory possesses an operator product expansion
in the sense of def.~\ref{opedef}, and suppose that the standard functionals $\sigma^{(i)}_{\M, x}$ in 
this operator product expansion satisfy (L), (M), and (A). Then the dependence of these standard 
functionals on the space and time orientations is expressed by the relation  
\begin{equation}
\label{key}
i^{F^{(i)}} (-1)^{M^{(i)}} \, \sigma_{\M, x}^{(i)}(\phi^{(j_1)}_\M(y_1) \cdots \phi^{(j_n)}_\M(y_n)) = 
i^F (-1)^M \, 
\sigma_{\overline \M, x}^{(i)}(\phi^{(j_n)}_{\overline \M}(y_n) \cdots \phi^{(j_1)}_{\overline \M}(y_1)),   
\end{equation}
for any finite number of local covariant fields, and any oriented and time oriented spacetime admitting
a spin structure, and all $y_j$ in some open neighborhood of $x$. 
Here, 
\begin{eqnarray}
F &=& F^{(j_1)} + \cdots + F^{(j_n)},\nonumber\\
M &=& M^{(j_1)} + \cdots + M^{(j_n)}, 
\end{eqnarray}
with $M^{(i)}$ the nunber of unprimed spinor indices of the field $\phi^{(i)}$, 
\begin{equation}
F^{(i)} = 
\begin{cases}
0 & \text{if $\phi^{(i)}$ is bosonic,}\\
1 & \text{if $\phi^{(i)}$ is fermionic,}
\end{cases}
\end{equation}
and it is understood that the map $\tilde I$ is used to identify the spinor 
indices corresponding to the space-time orientation $+o$  
on the left side with the spinor indices corresponding to the 
space-time orientation $-o$ on the right side of the above equation. 
\end{thm}

\paragraph{Remarks:} 
(1) The proof given below shows that relation~\eqref{key} holds true for any family of 
functionals $\sigma^{(i)}_{\M, x}$ with the properties (L), (M), (A), and eq.~\eqref{f}. The fact that
these functionals define an operator product expansion with property~\eqref{ope1} does not 
play any role in our proof. We also re-emphasize that it is neither 
assumed nor used anywhere in the proof
that the spin-statistics relation holds for the fields, i.e., it is not assumed 
that half odd-integer spin fields are fermionic and that integer spin fields are bosonic.

\medskip
\noindent
(2) It follows from condition~\eqref{f} on the functionals $\sigma^{(i)}$, that the number
$F$ is even respectively odd if and only if $\phi^{(i)}$ is bosonic respectively fermionic. 

\medskip
\noindent
(3) If the functionals $\sigma^{(i)}$ in the operator product expansion are 
real\footnote{We recall that when $\phi^{(i)}$ is a scalar field, then the functionals 
$\sigma^{(i)}_{\M, x} \in \Sigma(\M)$ are said to be 
real if $\overline{\sigma_{\M, x}^{(i)}(A)} = \sigma^{(i)}_{\M, x}(A^*)$ for all $A \in \cA(\M)$.
When the field $\phi^{(i)}_{\M}$ carries spinor indices, then, as explained above, 
$\sigma_{\M, x}^{(i)}$ should be viewed as an element in 
$\Sigma(\M) \otimes \cF_x(\M)$, with $\cF(\M)$ the vector bundle corresponding to the 
spinor character of the field. Such a functional is called real if it satisfies the same 
relation as in the scalar case, but where complex conjugation is now applied only to the first factor 
in the tensor product.}, one can reformulate the theorem as follows: If $\phi$ is a local covariant field
with $N$ primed and $M$ unprimed spinor indices, define a corresponding 
charge-conjugate field, $\phi^C$, by 
\begin{equation}
\phi^C(f)  = i^{F} (-1)^{M} \phi(\overline f)^*, 
\end{equation}
so that $\phi^C$ is now a local, covariant field that has $M$ primed and $N$ unprimed spinor indices.
Let $c^{(i)}_{\M, x}$ be the distributional coefficients appearing in the expansion of the 
product of the fields $\phi^{(j)}_\M$, $j=1, \dots, n$, on a spacetime $\M$ with a given space
and time orientation,  
\begin{equation}
\phi^{(1)}_{\M}(y_1) \dots \phi^{(n)}_{\M}(y_n) \sim \sum_{(i)} c^{(i)}_{\M, x}(y_1, \dots, y_n) \, \phi^{(i)}_\M(x)
\end{equation}
where ``$\sim$'' in the above relation is understood in the precise sense 
of def.~\ref{opedef}, and where the coefficients related to the standard functionals
in the theorem by eq.~\eqref{cidef}. Then eq.~\eqref{key} says that the distributional coefficients in the expansion of the product 
of the charge conjugate fields $\phi^{(j)C}_{\overline \M}(y_j)$ on the spacetime $\overline \M$ with the opposite space
and time orientation relative to that of $\M$ are given by the complex conjutates of the above coefficients $c^{(i)}_{\M, x}$
for the spacetime $\M$, i.e., 
\begin{equation}
\phi^{(1) C}_{\overline \M}(y_1) \dots \phi^{(n) C}_{\overline \M}(y_n) \sim \sum_{(i)} \overline{c^{(i)}_{\M, x}(y_1, \dots, y_n)} \, 
\phi^{(i) C}_{\overline \M}(x). 
\end{equation}
If these relations were honest equations rather than asymptotic relations as $(y_1, \dots, y_n) \to (x, \dots, x)$, then the 
$c^{(i)}$ could be viewed as structure constants of the algebra of fields, and the above relations
could be viewed as saying that the map sending any smeared local covariant field
$\phi_\M$ on $\M$ to its charge conjugate $\phi^{C}_{\overline \M}$ on $\overline \M$ defines an 
(anti-linear) isomorphism between the field algebra $\cA(\M)$ and the field algebra $\cA(\overline \M)$ of 
the spacetime associated with the opposite space and time orientation. In this sense, our theorem 
may be viewed as an analog of the PCT theorem in Minkowski spacetime. 

\paragraph{Proof of the theorem, scalar bosonic case:}
For simplicity, we will first treat the special case when all fields in the 
theory are scalar and bosonic. Then eq.~\eqref{key} reduces to 
\begin{equation}
\label{ci'}
\sigma_{\M, x}^{}(\phi^{(1)}_\M(y_1) \cdots \phi^{(n)}_\M(y_n)) = 
\sigma_{\overline \M, x}^{}(\phi^{(n)}_{\overline \M}(y_n) \cdots \phi^{(1)}_{\overline \M}(y_1)),   
\end{equation}
for all $y_j$ in some neighborhood $\N$ of $x$, where we have set $(j_k) = (k)$ without
loss of generality, and where we have dropped the 
superscript $(i)$ on $\sigma_{\M, x}$ to simplify the notation. 
Thus, when only scalar fields are present in the theory, 
our theorem will be proven if we can prove eq.~\eqref{ci'}.  

We will prove eq.~\eqref{ci'} using a particular family of metrics which
interpolates analytically between the metric $g_{ab}$ and the Minkowski metric. 
The construction of this family is as follows. In a convex normal 
neighborhood around the point $x$, we introduce Riemannian normal coordinates for our metric $g_{ab}$, 
denoted $y^\alpha = (y^0, y^1, y^2, y^3)$, so that the  the point $x$ has the coordinates $y^\alpha = 0$, 
and so that the coordinate components $g_{\mu\nu}$ of the 
metric satisfy $g_{\mu\nu}(0) = \eta_{\mu\nu}$. On this convex normal neighborhood 
of $x$, we define a family of metrics $g_{ab}^{(s)}$, $s \in I = (-1-c, 1+c), c>0$ via its
coordinate components by 
\begin{equation}
\label{gsdef}
g_{\mu\nu}^{(s)}(y^\alpha) =
\eta_{\mu\nu} + \sum_{n \ge 1} \frac{s^n}{n!} \sum_{\alpha_1 \dots \alpha_n}
y^{\alpha_1} \cdots y^{\alpha_n}
\frac{\partial^n g_{\mu\nu}(0)}{\partial y^{\alpha_1} \cdots \partial y^{\alpha_n}}.
\end{equation}
It is obvious from this expression that (in our convex normal neighborhood) $g_{ab}^{(0)}$ is the flat, Minkowskian 
metric, that $g^{(1)}_{ab}$ is equal to the original metric, and that $g_{ab}^{(s)}$ 
has an analytic dependence on the parameter $s \in I$. Since the statement of the theorem is completely 
local, we may pass from $\M$ to a neighborhood of $x$ on which all metrics
$g_{ab}^{(s)}$ are defined globally and which are globally hyperbolic 
with respect to these metrics for all $s \in I$. 
Furthermore, we will from now on view $\M$ as a 
neighborhood of the origin in $\mr^4$ by identifying a points $y \in \M$ with
their Riemannian normal coordinates, viewed as points in $\mr^4$.

If $(T, \epsilon_{abcd})$ is the time function respectively volume 
form defining the time and space orientation, $+o$,
of the spacetime $(\M, g_{ab}^{(1)})$, then it is clear that $\nabla_a T$ will 
remain timelike with respect to the metrics $g^{(s)}_{ab}$ in 
a neighborhood of the point $x$ for all $s \in I$. By shrinking $\M$ further if necessary, we 
can therefore assume without loss of generality that $T$ defines a corresponding time orientation 
on the spacetimes $(\M, g_{ab}^{(s)})$ for all $s \in I$. Similar remarks apply to the space orientation, as 
well as the reversed orientations $-o = (-T, \epsilon_{abcd})$. We have therefore defined 1-parameter
families of oriented and time oriented spacetimes 
\begin{equation}
\label{Msdef}
\M(s) = (\M, g_{ab}^{(s)}, +o), \quad \overline \M(s) = (\M, g_{ab}^{(s)}, -o).  
\end{equation}

By our assumption (A), we know that $\sigma_{\M, x}^{}$ are local and covariant 
functionals that vary analytically under analytic variations of the metric. 
This means by definition that we can pick a neighborhood $\N$ of the point $x$ such 
that the restriction to $\times^n \N$ of the family of $n$-point functions $\sigma_{\M(s), x}(\prod 
\phi^{(j)}_{\M(s)}(y_j))$ is a family of distributions which varies analytically with $s$ with respect to the 
cones $\Gamma_{\M(s)}$, and we may thus differentiate this family with respect to the parameter $s$ and 
set $s=0$ afterwards. An analogous statement holds of course for the family of spacetimes
$\overline \M(s)$ with the opposite orientations. If eq.~\eqref{ci'} holds for all spacetimes, 
then this gives
\begin{equation}
\label{cik1}
\frac{d^k}{d s^k} \, \sigma^{}_{\M(0), x}
(\phi^{(1)}_{\M(0)}(y_1) \cdots \phi^{(n)}_{\M(0)}(y_n)) = 
\frac{d^k}{d s^k} \, \sigma^{}_{\overline \M(0), x}(\phi^{(n)}_{\overline \M(0)}(y_n) \cdots \phi^{(1)}_{\overline \M(0)}(y_1)),  
\end{equation}
for all $(y_1, \dots, y_n) \in \times^n \N$, 
where $\M(0)$ is Minkowski spacetime $(\M, g_{ab}^{(0)})$ with space and time orientation $+o$, and where $\overline \M(0)$ is Minkowski
spacetime with the space and time orientation $-o$. This equation is therefore a necessary condition
for our theorem to be true. Our first mayor step in the proof is to show that it is also sufficient. 

It follows immediately from the microlocal spectrum condition together with the transformation 
rules of the analytic wave front set under diffeomorphims that
\begin{equation}
\WF_A(\sigma_{\overline \M, x}^{}(\phi^{(n)}_{\overline \M}(y_n) \cdots \phi^{(1)}_{\overline \M}(y_1)))
\subset \pi^* \Gamma_{\overline \M}, 
\end{equation}
where $\pi$ is the permutation 
\begin{equation}
\label{perm}
\pi = 
\left(
\begin{matrix}
1 & 2 & \dots & n\\
n & n-1 & \dots & 1
\end{matrix}
\right), 
\end{equation}
and where the set $\Gamma_{\overline \M}$ is defined as in~\eqref{gamtdef}, but with 
the orientations reversed relative to the original orientations, $o = (T, \epsilon_{abcd})$. 
We claim that $\pi^* \Gamma_{\overline \M} = \Gamma_\M$: Let $(y_1, k_1; \dots; y_n, k_n) \in 
\Gamma_\M$, which means that there exists a decorated graph $G(p)$ with $(p_e)^a \nabla_a T > 0$
such that $k_i = \sum_{e: i = s(e)} p_e - \sum_{e: i = t(e)} p_e$ for all $i$.  
(Remember that if $e$ is an edge joining $y_j$ and $y_k$ with $j<k$, then 
$j = s(e)$ and $k = t(e)$.) We need to show that $(y_n, k_n; \dots; y_1, k_1) \in 
\Gamma_{\overline \M}$. Consider the graph $\bar G(p)$ whose edges and vertices are
identical to the edges and vertices of $G(p)$, but which are decorated with the 
covectors $-p_e$, which are future pointing with respect to the time function
$-T$ on $\overline \M$. Note that the notion of source $\bar s(e)$ and target $\bar t(e)$ 
relative to the ordering $(y_n, \dots, y_1)$ of the vertices is opposite
to the above notion of source and target relative to the ordering $(y_1, \dots, y_n)$,
so that if $e$ is an edge in $\bar G(p)$ joining $y_j$ and $y_k$ with $j<k$, then 
$j = \bar t(e)$ and $k = \bar s(e)$ relative to the ordering $(y_n, \dots, y_1)$. 
It is a trivial consequence of these definitions that $k_i$ can be written alternatively
as $k_i = \sum_{e: i = \bar s(e)} (-p_e) - \sum_{e: i = \bar t(e)} (-p_e)$ for all $i$, which
displays $(y_n, k_n; \dots; y_1, k_1)$ as the element in $\Gamma_{\overline \M}$ associated
with the graph $\bar G(p)$.

Let $u$ be the distribution on $\times^n \M$ given by
the difference between the left and right side of eq.~\eqref{ci'}. Then $u$ is the difference 
of two distributins whose analytic wave front set is contained in $\Gamma_\M$. 
Hence, by the rules for calculating the analytic wave front set of sums of 
distributions, we have $\WF_A(u) \subset \Gamma_\M$. By a similar argument, if
$u^{(s)}$ is the family of distributions defined as the difference of the left
minus right side of eq.~\eqref{ci'} with $\M$ and $\overline \M$ replaced by 
$\M(s)$ and $\overline \M(s)$, then $u^{(s)}$ is an analytic family of distributions
on $\times^n \N$ relative to the conic sets $\Gamma_{\M(s)}$. Furthermore, since 
we assume that eq.~\eqref{cik1} holds, we have that 
\begin{equation}
\frac{d^k}{ds^k} u^{(s)}(y_1, \dots, y_n) |_{s=0} = 0 \quad \forall k, (y_1, \dots, y_n) \in \times^n \N.
\end{equation}
We need to show that it follows that $u^{(s)}(y_1, \dots, y_n) = 0$ for all $s \in I$ and $(y_1, \dots, y_n)
\in \times^n \N$. Since we have identified $\M$ with a neighborhood of in $\mr^4$
and the point $x$ with the origin via Riemannian normal coordinates, we 
may take the neighborhood $\N$ to be the ball $B_r$ of radius $r$ around the origin 
in $\mr^4$ with respect to the Euclidean norm 
\begin{equation}
\label{norm}
\|y\| = \sqrt{|y^0|^2 + |y^1|^2 +|y^2|^2 +|y^3|^2},   
\end{equation}
which depends on our choice of coordinates. It is known~\cite[lem. 4.2]{bfk} that each component
$(\Gamma_\M)_{(y_1, \dots, y_n)}$, of $\Gamma_\M$ in the cotangent space 
$T^*_{(y_1, \dots, y_n)}(\times^n \M) \setminus \{0\}$, 
is a proper\footnote{A cone is said to be proper if it does not 
contain any straight line.}, 
closed convex cone, which we identify with 
a proper, closed convex cone in $\mr^{4n} \setminus \{0\}$. Moreover, it can be seen that 
$r>0$ can be chosen so small that 
\begin{equation}
\bigcup_{(y_1, \dots, y_n) \in \times^n B_r, s \in I} (\Gamma_{\M(s)})_{(y_1, \dots, y_n)} \subset C
\end{equation}
where $C$ is a proper, closed, convex cone in $\mr^{4n}$. 
Our claim that $u^{(s)}(y_1, \dots, y_n) = 0$ now 
follows from the following lemma, which we shall prove in appendix C.

\begin{lemma}
\label{appclem}
Let $u^{(s)} \in \cD'(X)$, $X$ an open subset of $\mr^m$, a 
family of distributions that depends analytically on $s \in I = (a, b)$ with respect to 
conic sets $K^{(s)} \subset X \times (C \setminus \{0\})$, where $C$ is 
a closed, proper convex cone in $\mr^m$. Suppose that
\begin{equation}
\frac{d^k}{d s^k} u^{(s)} |_{s=s_0} = 0
\end{equation} 
for all $k$ and some $s_0 \in I$. Then $u^{(s)} = 0$ for all $s \in I$. 
\end{lemma}

Thus, we have obtained the important intermediate result that 
our theorem will be proven if eq.~\eqref{cik1} can be shown. 

\medskip

It follows immediately from the definition of our analytic family of 
metrics that the map $\chi: y \to -y$ satisfies
\begin{equation}
\chi^* g^{(s)}_{ab} = g^{(-s)}_{ab}.  
\end{equation}
Moreover, it is clear that $\chi$ reverses the space and time 
orientation in the sense that $\chi^* o = (\chi^* T, \chi^* \epsilon_{abcd})$ defines
the same space and time orientation as $-o = (-T, \epsilon_{abcd})$. This 
shows that $\chi$ is an orientation {\em preserving} isometry between the 
space and time orientated spacetimes $\M(s)$ and  $\overline \M(-s)$ given in eq.~\eqref{Msdef}. 
By the locality and covariance property of the fields, we therefore have that
\begin{equation}
\chi^* \phi_{\overline \M(-s)}(y) 
= \phi_{\overline \M(-s)}(-y) = \alpha_\chi^{} (
\phi_{\M(s)}(y) ), 
\end{equation}
and by the locality and covariance of the standard functionals, we have that 
\begin{equation}
\sigma^{}_{\overline \M(-s), x} 
\circ \alpha_\chi^{} \xsim \sigma^{}_{\M(s), x}.  
\end{equation} 
Putting this together, we therefore know that 
\begin{equation}
\sigma^{}_{\M(s), x}
(\phi^{(1)}_{\M(s)}(y_1) \cdots \phi^{(n)}_{\M(s)}(y_n)) = 
\sigma^{}_{\overline \M(-s), x}(\phi^{(1)}_{\overline \M(-s)}(-y_1) \cdots \phi^{(n)}_{\overline \M(-s)}(-y_n)),   
\end{equation}  
for all $(y_1, \dots, y_n) \in \times^n B_r$, all $s \in I$, for some $r>0$.
If we now differentiate both sides of this equation $k$ times with respect to $s$ at $s=0$, and substitute
the result into eq.~\eqref{cik1}, we get the important intermediate 
result that the theorem will be proven if we can show 
\begin{equation}
\label{cik}
\frac{d^k}{d s^k} \, \sigma^{}_{\M(0), x}
(\phi^{(1)}_{\M(0)}(y_1) \cdots \phi^{(n)}_{\M(0)}(y_n)) = (-1)^k
\frac{d^k}{ds^k} \, \sigma^{}_{\M(0), x}(\phi^{(n)}_{\M(0)}(-y_n) \cdots \phi^{(1)}_{\M(0)}(-y_1)),  
\end{equation}
for all $k$, $(y_1, \dots, y_n) \in \times^n B_r$ and some $r>0$, 
where $\M(0)$ is Minkowski spacetime $(\M, g_{ab}^{(0)})$ with the orientation $+o$. Thus, by the 
preceeding steps, we have managed to transform the original problem of 
proving identity~\eqref{ci'} between distributions associated with 
spacetimes $(\M, g_{ab}^{(1)})$ with space and time orientations $+o$ respectively $-o$, 
to the problem of proving relations~\eqref{cik} for a set of distributions
on Minkowski spacetime $(\M, g_{ab}^{(0)})$ with a single orientation, $+o$.
The remainder of this proof therefore consists of showing that 
these relations are indeed true.

\medskip

For this, we need to analyze the $s$-derivatives of the distributions~\eqref{sfunc} for our 
particular family of spacetime metrics~\eqref{gsdef} and orientation $+o$. 
Such an analysis was carried out in similar context in~\cite[thm. 4.1]{hw2} in order to 
derive a ``scaling expansion'' for certain distributions that arise in the 
context of perturbative interacting quantum field theories in curved spacetimes. The 
properties of the distributions considered in~\cite{hw2} which enter the analysis
are (a) that they are locally and covariantly constructed from the metric near a reference point, $x$, (b) that they 
depend analytically on the metric in analytic spacetimes, (c) that they depend 
smoothly on the metric in smooth spacetimes, and (d) that they have a certain scaling 
behavior under rescalings of the metric by constant conformal factor. Using only these 
properties, it was shown that the $k$-th derivative with respect to $s$ of the 
family of these distributions corresponding to the spacetime metrics defined in~\eqref{gsdef}
can be written as a linear combination of curvature terms of the appropriate ``dimension'', 
times Lorentz-invariant Minkowski space distributions (they also satisfy other properties, 
but these are not relevant in the present context). Inspection of the proof of 
this statement given in~\cite{hw2} shows that in analytic spacetimes, it only relies on (a) and (b) above, 
but not on (c) and (d). Furthermore, one easily sees that the arguments given in ~\cite{hw2} will still be valid
when properties (a) and (b) are replaced by the essentially identical properties (L) and (A) assumed 
for our distributions~\eqref{cidef}.
(In fact, the precise form of our conditions (L) and (A) has been chosen precisely so that the 
arguments of~\cite{hw2} are still valid.) We therefore conclude 
that the expression on the left side of eq.~\eqref{cik} can be decomposed into a sum of curvature terms of the appropriate
dimension, times Lorentz invariant distributions in Minkowski spacetime. However, 
since assumtions (L) and (A) are weaker than the requirements (a) and (b) used 
in~\cite[thm. 4.1]{hw2} in that they hold only for an arbitrary small neighborhood of the reference
point, $x$, one gets only the weaker result 
that the Minkowski space distributions are in fact only defined in 
some neighborhood of the origin (which we take to be a ball), and that they are invariant
only under those Lorentz transformations $\Lambda$ that are sufficiently close to the 
identity. More precisely, there  exists an $r > 0$ such that 
\begin{equation}
\label{scexp}
\frac{d^k}{d s^k} \, \sigma^{}_{\M(0), x}
(\phi^{(1)}_{\M(0)}(y_1) \cdots \phi^{(n)}_{\M(0)}(y_n)) = 
\sum_j \sum_{\mu_1 \dots \mu_j} C^{(J)\mu_1 \dots \mu_j}(x) \, W_{\mu_1 \dots \mu_j}^{(J)}(y_1, \dots, y_n)
\end{equation}
for all $(y_1, \dots, y_n) \in \times^n B_r$, where
$W^{(J)}$ are tensor-valued distributions on $\times^n B_r$ and 
where $(J) = (12 \dots n)$ is a shorthand for the indices labelling the fields. 
The expressions $C^{(J)}$ are the coordinate components in Riemannian normal coordinates of curvature tensors 
that are polynomials  
\begin{equation}
C^{(J)m_1 \dots m_j}  (g_{ab}(x), R_{abcd}(x), \dots,
\nabla_{(e_1} \cdots \nabla_{e_{k-2})} R_{abcd}(x))
\end{equation}
of the metric, the Riemann tensor and its covariant derivatives at $x$. 
Each monomial in $C^{(J)}$ contains precisely $k$ derivatives of the metric, implying that
\begin{equation}
j = k \,\, {\rm mod} \,\, 2.
\end{equation}
The $W^{(J)}$ have the further property:
\begin{enumerate}
\item[(i)]
There exists an $r, \delta > 0$ such that 
\begin{equation}
W^{(J)}(\Lambda y_1, \dots, \Lambda y_n) = 
D(\Lambda) W^{(J)}(y_1, \dots, y_n)
\end{equation}
for all $\Lambda \in \cL$ with $\| \Lambda - 1\| < \delta$ 
and all $(y_1, \dots, y_n) \in \times^n B_{r-\delta}$, where the norm of a linear transformation 
is defined using the Euclidean norm~\eqref{norm}, and where $D(\Lambda)$ is the tensor representation
\begin{equation}
D(\Lambda)^{\nu_1 \dots \nu_j}_{\mu_1 \dots \mu_j} =
\Lambda^{\nu_1}_{\mu_1} \cdots \Lambda^{\nu_j}_{\mu_j}. 
\end{equation}  
\end{enumerate}
Furthermore, since the restriction of $\sigma^{}_{\M(s), x}$ to a sufficiently small neighborhood of 
$x$ satisfies the microlocal spectrum condition for the cone $\Gamma_{\M(s)}$, and since this family 
has an analytic dependence on $s$ with respect to these cones, we have, 
by the same arguments as in the remark at the end of section~4 of~\cite{hw2},
that  
\begin{enumerate}
\item[(ii)]
There exists an $r > 0$ such 
that the restriction of $W^{(J)}$ to $\times^n B_r$ has 
analytic wave front set
\begin{equation}
\WF_A(W^{(J)}) \subset \Gamma_{\M(0)}. 
\end{equation}
Here, $\Gamma_{\M(0)}$ is the cone~\eqref{gamtdef} defined 
with respect to the Minkowskian metric $g_{ab}^{(0)}$ and orientation $+o$. 
Using our Riemannian normal coordinates $(y^0, y^1, y^2, y^3)$ to identify 
$\M$ with a subset of $B_r$ and assuming that 
$\nabla_a y^0$ is future pointing with respect to $+o$, it can be written as 
\begin{eqnarray}
\label{uawfs}
\Gamma_{\M(0)} &=&
\Big\{(y_1, k_1; \dots; y_n, k_n) 
\in T^*(\times^n B_r) \setminus \{0\} \,\, \Big| \,\, 
\text{$\exists p_{ij} \in \bar V_+$, $n \ge j > i \ge 1$:} \nonumber\\
&& \text{
$k_i = \sum_{j:j>i} p_{ij} - \sum_{j:j<i} p_{ji}$ for all $i$}
\Big\}, 
\end{eqnarray}
where $\bar V^+$ is the closure of the forward light cone $V^+$ in Minkowski space, 
\begin{equation}
V^\pm = \{ k \in \mr^4 \mid \eta_{\mu\nu} k^\mu k^\nu > 0, \quad \pm k^0 > 0\}.
\end{equation}
\end{enumerate}

Besides the above properties (i) and (ii) for the $W^{(J)}$ in the 
expansion~\eqref{scexp}, we will now derive one more property, (iii),
from the fact that the coefficients in our operator product expansion 
are not just arbitrary local covariant distributions with a specific analytic dependence on the metric, but arise in 
fact from a set of linear functionals 
on the algebras of observables $\cA(\M)$. To exploit this fact, we consider the multilinear
maps on $\times^n \cD(\M)$ defined by 
\begin{equation}
\label{phidef}
(f_1, \dots, f_n) \to \sigma^{}_{\M, x}
( \phi^{(1)}_{\M}(f_1) \cdots 
[\phi^{(k)}_{\M}(f_k), \phi^{(k+1)}_{\M}(f_{k+1})] \cdots
\phi^{(n)}_{\M}(f_n) ).
\end{equation} 
Then by eq.~\eqref{com}, the right side of the above equation
will vanish if the supports of $f_k$ and $f_{k+1}$ are spacelike
related in $\M$ with respect to the metric $g_{ab}^{(1)}$. Consider now the analytic
family of metrics $g^{(s)}_{ab}$ constructed above
in eq.~\eqref{gsdef} and a set of testfunctions on $\M$
such that the supports of $f_k$ and $f_{k+1}$ are spacelike related with 
respect to the flat Minkowskian metric $g_{ab}^{(0)}$. Then the supports of 
$f_k$ and $f_{k+1}$ will continue to be spacelike related also
with respect to the metrics $g_{ab}^{(s)}$ for sufficiently small $s$, so that 
\begin{equation}
\label{phidef'}
\sigma^{}_{\M(s), x}( \phi^{(1)}_{\M(s)}(f_1) \cdots 
[\phi^{(k)}_{\M(s)}(f_k), \phi^{(k+1)}_{\M(s)}(f_{k+1})] \cdots
\phi^{(n)}_{\M(s)}(f_n)) = 0
\end{equation} 
will hold provided that $s$ is sufficiently small. Since the functionals $\sigma_{\M, x}$ depend analytically
on the metric, it is possible to find an open neighborhood of $x$ in $\M$ such that the restriction of 
eq.~\eqref{phidef'} to testfunctions supported in that neighborhood (which we shall again denote $\M$)
defines a distribution that depends analytically on $s$ and that vanishes for sufficiently small $s$
when the supports of $f_k$ and $f_{k+1}$ are spacelike related with respect to the Minkowskian metric
$g_{ab}^{(0)}$. We therefore find that  
\begin{equation*}
(f_1, \dots, f_n) \to \frac{d^k}{ds^k} \, \sigma^{}_{\M(s), x}
( \phi^{(1)}_{\M(s)}(f_1) \cdots 
[\phi^{(k)}_{\M(s)}(f_k), \phi^{(k+1)}_{\M(s)}(f_{k+1})] \cdots
\phi^{(n)}_{\M(s)}(f_n) )|_{s=0}
\end{equation*} 
is a distribution on $\times^n \cD(\M)$ that vanishes whenever the 
supports of $f_k$ and $f_{k+1}$ are spacelike separated with respect to 
the Minkowskian metric $g_{ab}^{(0)}$. Since the distributions $W^{(J)}$ are related to the 
above derivatives via the expansion~\eqref{scexp}, we get from this:
\begin{enumerate}
\item[(iii)]
There exists an $r > 0$ such that if $\|y_j\| < r$ and $y_i$ and $y_{i+1}$ 
are spacelike to each other with respect to the Minkowski metric $\eta_{\mu\nu}$, 
then there holds  
\begin{equation}
W^{(J)}(y_1, \dots, y_i, y_{i+1}, \dots, y_n) = 
W^{(\pi_{i,i+1} J)}(y_1, \dots, y_{i+1}, y_i, \dots, y_n),  
\end{equation}
where $(\pi_{k,k+1} J)$ stands for $(1 \dots (k+1)k \dots n)$.
\end{enumerate}

We have argued so far that the theorem will be proven if we can show eq.~\eqref{cik}. 
If we now substitute the expansion~\eqref{scexp} into eq.~\eqref{cik}, and use that $j = k$
modulo 2, we see eq.~\eqref{cik} will follow if we can show that there is a $r > 0$ 
such that  
\begin{equation}
\label{WCPT}
W^{(J)}(y_1, \dots, y_{n}) 
= (-1)^j W^{(\pi J)}(-y_n, \dots, -y_1) 
\end{equation}
for all $y_j \in B_r$ in the sense of distributions, where $\pi$ is the 
permutation~\eqref{perm}. Since we already know that the 
distributions $W^{(J)}$ satisfy properties (i), (ii) and (iii) above, the proof of the 
theorem will therefore be complete once we have shown the following proposition:

\begin{prop}
Suppose that $W^{(J)}_{\mu_1 \dots \mu_j}$ are tensor-valued 
distributions on $\times^n B_r$ for which there holds (i), (ii) and (iii). Then 
there is some $r > 0$ such that $W^{(J)}$ satisfies eq.~\eqref{WCPT} 
within $\times^n B_r$. 
\end{prop}

\begin{proof}
Consider now the linear transformation $f: (\xi_1, \dots, \xi_n) \to (y_1, \dots, y_n)$ on $\mr^{4n}$
defined by 
\begin{equation}
\label{Ldef}
f: y_i = \xi_i + \xi_{i+1} + \cdots + \xi_n \quad \text{for all $i$,}
\end{equation}
so that $\xi_n = y_n$ and $\xi_i = y_i - y_{i+1}$ for all $i \neq n$. For $(\xi_1, \dots, \xi_n) 
\in \times^n B_r$, $r > 0$ sufficiently small, define
$\cW^{(J)}(\xi_1, \dots, \xi_n) = W^{(J)}(f(\xi_1, \dots, \xi_n))$, which expresses 
$W^{(J)}$ in terms of relative coordinates about the ``center of mass'' point $\xi_n = y_n$.
By the rules for calculating the analytic wave front set of the pull back of a distribution under 
an analytic map, we have 
\begin{eqnarray}
&&\WF_A(\cW^{(J)}) = f^* \WF_A(W^{(J)}) \nonumber\\
&=& \Big\{(\xi_1, \ell_1; \dots; \xi_n, \ell_n) \in T^*(\times^n B_r) \setminus \{0\} \,\,\, \Big| \,\, 
\exists (y_1, k_1; \dots; y_n, k_n) \in \WF_A(W^{(J)}): \nonumber \\
&&\ell_i = k_1 + \cdots + k_i, \quad
y_i = \xi_i + \xi_{i+1} + \cdots + \xi_n \Big\}.
\end{eqnarray}
By (ii), we know that $(y_1, k_1; \dots; y_n, k_n)$ is in $\WF_A(W^{(J)})$ if and only if there 
exists a set of covectors $p_{ij} \in \bar V^+, i < j$ such that $k_i = \sum_{j: j<i} p_{ji} - \sum_{j: j>i} p_{ij}$
for all $i$. Thus, if $(\xi_1, \ell_1; \dots; \xi_n, \ell_n)$ is in $\WF_A(\cW^{(J)})$, then we must have  
\begin{eqnarray}
\ell_i = \sum_{j=1}^i k_j &=& \sum_{j=1}^i \left( \sum_{l: l>j} p_{jl} - \sum_{l: l<j} p_{lj} \right)\\
&=& \sum_{j: j > i} (p_{1j} + p_{2j} + \cdots + p_{ij}),   
\end{eqnarray}
where the equality in the second line can be proved by induction in $i$.
Thus, $\ell_i \in \bar V_+$ for all $i$, and in particular $\ell_n = 0$. We have thus shown that 
\begin{equation}
\label{wfw'}
\WF_A(\cW^{(J)}) \subset (\times^{n} B_r) \times (\bar V^+ \times \cdots \times \bar V^+ \times \{0\}).
\end{equation}  
We will now use this information about the analytic wave front set of $\cW^{(J)}$ to show 
that it is the boundary value of some analytic function. This will follow from the 
the following key result about distributions whose
analytic wave front set is contained in the dual of an open, convex cone~\cite[thm. 8.4.15]{h}:
\begin{thm}
\label{bvthm}
Let $u$ be a distribution on $X \subset \mr^m$ with $\WF_A(u) \subset X \times K^D$, where\footnote{
Our definition of the dual cone differs trivially from that employed in~\cite[thm. 8.4.15]{h} since our
convention for the Fourier transform is opposite to the convention employed in~\cite{h}.}
\begin{equation}
K^D = \{ k \in \mr^m \mid k \cdot x \le 0 \quad \forall x \in K \}
\end{equation} 
is the dual of an open convex cone $K \subset \mr^m$, with $k \cdot x$ the standard 
inner product in $\mr^m$. If $X_0 \subset X$ is an open 
subset with compact closure $\bar  X_0 \subset X$, then one can find a $\gamma > 0$ and a function $U$ analytic 
in $\{x + iy \in \mc^m \mid x \in X_0, y \in K, \|y\| < \gamma \}$ 
such that $u$ is the boundary value of $U$, 
\begin{equation}
u(f) = \lim_{y \in K, y \to 0} \int U(x + iy) f(x) \, d^m x
\end{equation}
which we write as 
\begin{equation}
u(x) = \bv_{y \in K, y \to 0} U(x + iy).
\end{equation}
\end{thm}
The closed forward lightcone is the dual of the open past lightcone, $\bar V^+ = (V^-)^D$, therefore
\begin{equation}
\bar V^+ \times \cdots \times \bar V^+ \times \{0\}
= (V^- \times \cdots \times V^- \times \mr^4)^D \subset \mr^{4n}, 
\end{equation}
so eq.~\eqref{wfw'} tells us 
in combination with the above theorem that $\cW^{(J)}$ is the boundary value 
\begin{equation}
\label{bv}
\cW^{(J)}(\xi_1, \dots, \xi_n) = 
\bv_{\eta_j \to 0, \eta_j \in V^- \,\, \forall j \neq n} \cW^{(J)}
(\xi_1 + i\eta_1, \dots, \xi_{n} + i \eta_{n})
\end{equation}
of a function $\cW^{(J)}(\zeta_1, \dots, \zeta_n)$ that is holomorphic in the domain
\begin{equation}
\label{dom}
\mathcal{T}_n = \{(\zeta_1, \dots, \zeta_n) \in \mc^{4n} \mid 
\text{$\| \zeta_j \| < r$ for all $j$, $\Im \zeta_j \in V^-$ for all $j \neq n$}\} 
\end{equation}
for some $r >0$.
Thus, we have shown that property (ii) of the distributions $W^{(J)}$ implies that 
$\cW^{(J)}$ is the boundary value of a function that is holomorphic on $\mathcal{T}_n$. 

We next want to show that the analytic continuations $\cW^{(J)}(\zeta_1, \dots, \zeta_n)$
are Lorentz invariant. For this let $\Lambda \in \cL$ with $\|\Lambda - 1\| < \delta$, 
and consider the function
\begin{equation}
(\zeta_1, \dots, \zeta_n) \to 
\cW^{(J)}(\Lambda \zeta_1, \dots, \Lambda \zeta_n) - 
D(\Lambda) \cW^{(J)}(\zeta_1, \dots, \zeta_n)
\end{equation}
for $(\zeta_1, \dots, \zeta_n) \in \mathcal{T}_n$ such that $\| \zeta_j \| < r - \delta$ for all $j$. 
The boundary value of this function as $\Im \zeta_j \to 0$ vanishes by (i). Therefore, by 
the ``edge-of-the-wedge theorem'' (see e.g.~\cite[thm. 2.17]{sw}), this function itself has to vanish, 
\begin{equation}
\label{lor'}
\cW^{(J)}(\Lambda \zeta_1, \dots, \Lambda \zeta_n) = 
D(\Lambda) \cW^{(J)}(\zeta_1, \dots, \zeta_n)
\end{equation}
for all $\Lambda \in \cL$ such that $\|\Lambda - 1\| < \delta$ and such that $(\zeta_1, \dots, \zeta_n)$ and
$(\Lambda \zeta_1, \dots, \Lambda \zeta_n)$ are in $\mathcal{T}_n$. 

We finally would like to use 
property (iii) to infer a corresponding property for $\cW^{(J)}$. Let $(\xi_1, \dots, \xi_n) \in 
\times^n B_r$, $(y_1, \dots, y_n) = f(\xi_1, \dots, \xi_n)$ such that all the difference
vectors $y_i - y_j$ are spacelike related with respect to $\eta_{\mu\nu}$. Then
since $\cW^{(J)} =f^* W^{(J)}$, and since $W^{(J)}$ satisfies (iii), we conclude that there 
exists an $r > 0$ such that
\begin{equation}
\label{com'}
\cW^{(J)}(\xi_1, \dots, \xi_{n-1}, \xi_n) = 
\cW^{(\pi J)}(-\xi_{n-1}, \dots, -\xi_{1}, \sum_{i=1}^{n} \xi_i),  
\end{equation}
in the sense of distributions. 

We have thus altogether shown that properties (i), (ii) and (iii) for $W^{(J)}$ imply that 
$\cW^{(J)} = f^* W^{(J)}$, with $f$ given by~\eqref{Ldef}, is the boundary value of an analyic function 
$\cW^{(J)}(\zeta_1, \dots, \zeta_n)$ on the domain $\mathcal{T}_n$, satisfying eqs.~\eqref{lor'} and~\eqref{com'}. When 
expressed in terms of $\cW^{(J)}$, the assertion~\eqref{WCPT} of the proposition reads
\begin{equation}
\label{WPCT'}
\cW^{(J)}(\xi_1, \dots, \xi_{n-1}, \xi_n) = (-1)^j
\cW^{(\pi J)}(\xi_{n-1}, \dots, \xi_{1}, -\sum_{i=1}^{n} \xi_i),  
\end{equation}
in the sense of distributions on $\times^n B_r$ for some $r > 0$. We have therefore reached
the important intermediate conclusion that the proposition will be shown if we can show 
that eq.~\eqref{WPCT'} holds for any $\cW^{(J)}$ which is the boundary value of an analytic function 
on the domain $\mathcal{T}_n$ satisfying eqs.~\eqref{lor'} and~\eqref{com'}. 

One notes that 
these properties of the functions $\cW^{(J)}$ resemble properties of the Wightman functions
in Minkowski spacetime~\cite{sw} (when expressed in relative coordinates), 
and that relation eq.~\eqref{WPCT'} likewise resembles a property of the Wightman functions 
that reflects the PCT invariance of a Wightman field theory, and our proof of eq.~\eqref{WPCT'}
will indeed follow closely the proof of the PCT theorem in Minkowski spacetime, see especially~\cite{sw}. One also notes 
that there are, however, two important differences between our functions $\cW^{(J)}$ and the Wightman functions (expressed
in relative coordinates) in Minkowski spacetime. Firstly, our functions $W^{(J)}$ 
are by contrast with the Wightman functions {\em not} translation invariant, so the 
relations~\eqref{com'} reflecting the local commutativity are not identical to the 
corresponding relations for the Wightman functions. Secondly, our distributions $\cW^{(J)}$ 
as well as their analytic extensions are defined only locally and the Lorentz invariance
eq.~\eqref{lor'} holds only locally. On the other hand, one finds that global translation invariance
and invariance under global Lorentz transformations play an important role once one looks
at the details of the proof (see e.g.~\cite{sw}) of the PCT theorem in Minkowski space. For these reasons, the arguments
given e.g. in~\cite{sw} cannot be taken over wholesale, but must be carefully adapted. 

\medskip

First, we notice that the transformation law eq.~\eqref{lor'} of the $\cW^{(J)}$ 
not only holds for proper Lorentz transformations $\Lambda$ such that $\|\Lambda -1\| < \delta$
and such that $(\zeta_1, \dots, \zeta_n)$ and $(\Lambda \zeta_1, \dots, \Lambda \zeta_n)$ 
are in $\mathcal{T}_n$, but moreover for any 
rotation of the form
\begin{equation}
\label{Rdef}
R(\varphi) = 
\left(
\begin{matrix}
1 & 0 & 0 & 0 \\
0 & \cos \varphi & \sin \varphi & 0 \\
0 & -\sin \varphi & \cos \varphi & 0 \\
0 & 0 & 0 & 1 
\end{matrix}
\right). 
\end{equation}
This can easily be proven by noting that such a rotation leaves the region $\mathcal{T}_n$ invariant
and that it can be written as a product of $N$ rotations $R(\varphi/N)$, each of which 
satisfy $\|R(\varphi/N) - 1\| < \delta$. The invariance then follows by applying the 
transformation rule~\eqref{lor'} to each such small rotation in turn and using the 
group character of the transformation rule.   

We will now show that the transformation law can be further generalized to more general transformations
by invoking the analyticity of the functions $\cW^{(J)}$. Consider the abelian group of complex 
Lorentz transformations
\begin{equation}
\Lambda(\alpha + i\beta) =
\left(
\begin{matrix}
\sinh(\alpha + i\beta) & 0 & 0 & \cosh(\alpha + i\beta) \\
0 & 1 & 0 & 0 \\
0 & 0 & 1 & 0 \\
\cosh(\alpha + i\beta) & 0 & 0 & \sinh(\alpha + i\beta) 
\end{matrix}
\right),  \quad \alpha, \beta \in \mr, 
\label{boost}
\end{equation}
which corresponds to a real, proper orthochronous Lorentz transformation if 
$\beta = 0$. The action of such a transformation on a complex vector, 
$\zeta \to \hat \zeta = \Lambda(\alpha + i\beta)\zeta$, can be written as
\begin{equation}
\label{lab}
\hat \zeta^+ = \e^{\alpha + i\beta} \zeta^+, 
\quad \hat \zeta^- = \e^{-\alpha - i\beta} \zeta^-, \quad \hat \zeta^1 = \zeta^1, \quad
\hat \zeta^2 = \zeta^2, 
\end{equation}  
where we have introduced the notation $\zeta^\pm = (\zeta^0 \pm \zeta^3)/\sqrt{2}$ for every 
complex four vector. We now have (compare~\cite[thm. 2.11]{sw}):
\begin{lemma}
\label{extension}
The functions $\cW^{(J)}(\zeta_1, \dots, \zeta_n)$ possess a unique, 
single-valued analytic continuation to the extended tube domain
\begin{equation}
\mathcal{T}_n' = \{ (\Lambda(\alpha + i\beta)\zeta_1, \dots, \Lambda(\alpha + i\beta)\zeta_n) 
\mid 
(\zeta_1, \dots, \zeta_n) \in \mathcal{T}_n, \alpha, \beta \in \mr\} \subset \mc^{4n}, 
\end{equation}
which transforms as 
\begin{equation}
\label{wtrans}
\cW^{(J)}(\Lambda(\alpha + i\beta) \zeta_1, \dots, \Lambda(\alpha + i\beta) \zeta_n) = 
D(\Lambda(\alpha + i\beta)) \cW^{(J)}(\zeta_1, \dots, \zeta_n)
\end{equation}
for all $(\zeta_1, \dots, \zeta_n) \in \mathcal{T}_n'$ and all $\alpha, \beta$. 
\end{lemma}
\begin{proof}
We already know by 
eq.~\eqref{lor'} that eq.~\eqref{wtrans} holds if $\beta=0$, if $(\zeta_1, \dots, \zeta_n)$ as well as 
$(\Lambda(\alpha) \zeta_1, \dots, \Lambda(\alpha) \zeta_n)$ are in $\mathcal{T}_n$
and if $\alpha$ is in a sufficiently small real neighborhood of $0$. Since $\Lambda(\alpha)$ is real analytic in $\alpha$, 
both sides of eq.~\eqref{lor'} define analytic functions of $\alpha$ in this real neighborhood of 0. 
Therefore eq.~\eqref{wtrans} holds for all $\alpha + i\beta$ in a sufficiently small complex neighborhood of 0
such that $(\Lambda(\alpha + i\beta) \zeta_1, \dots, \Lambda(\alpha + i\beta) \zeta_n) \in \mathcal{T}_n$. 
If $(\zeta_1, \dots, \zeta_n)$ is in $\mathcal{T}_n$ but 
$(\Lambda(\alpha + i\beta) \zeta_1, \dots, \Lambda(\alpha + i\beta)
\zeta_n)$ is not in ${\mathcal T}_n$, then the right side of eq.~\eqref{wtrans} is initially not defined and we try to define it by the 
left side in this case. It may happen that a point $(\xi_1, \dots, \xi_n) \in \mathcal{T}_n'$ can be reached in different ways from 
elements in $\mathcal{T}_n$, i.e., that it can be written as $(\Lambda(\alpha_1 + i\beta_1) \zeta_1, \dots, \Lambda(\alpha_1 + i\beta_2)\zeta_n)$
or $(\Lambda(\alpha_2 + i\beta_2) \rho_1, \dots, \Lambda(\alpha_2 + i\beta_2)\rho_n)$ where $(\zeta_1, \dots, \zeta_n)$ and $(\rho_1, 
\dots, \rho_n)$ are in $\mathcal{T}_n$. Unless these different ways of writing $(\xi_1, \dots, \xi_n)
\in \mathcal{T}_n'$ give rise to the same definition of $\cW^{(J)}(\xi_1, \dots, \xi_n)$, our proposed extension of $\cW^{(J)}$ will not be 
single valued. Thus, the nontrivial task is to show that eq.~\eqref{wtrans} holds when $(\zeta_1, \dots, \zeta_n)$ and  
$(\Lambda(\alpha + i\beta) \zeta_j, \dots, \Lambda(\alpha + i\beta) \zeta_n)$ 
are in $\mathcal{T}_n$, where $\alpha = \alpha_1 - \alpha_2$ and 
$\beta = \beta_1 - \beta_2$. We already know that eq.~\eqref{wtrans} holds when $\alpha + i\beta$ is sufficiently close 
to 0. Therefore, by the well known method of analytic continuation by overlapping neighborhoods combined with the 
group character of the transformation law eq.~\eqref{wtrans}, this equation will follow if we can show that there exists 
a continuous curve $t \to \gamma(t) = \alpha(t) + i\beta(t)$ such that $\gamma(0) = 0, \gamma(1) = \alpha + i\beta$ and 
such that $(\Lambda(\gamma(t))\zeta_1, \dots, \Lambda(\gamma(t)) \zeta_n) \in \mathcal{T}_n$ for all $0 \le t \le 1$. 

Thus, our construction of the analytic extension of $\cW^{(J)}$ to the extended domain $\mathcal{T}_n'$ will be complete if
we can construct such a curve $\gamma$. Without loss of generality we can assume that $0 \le \beta \le \pi$. 
Our proposal for the curve $\gamma$ is then 
\begin{equation}
\gamma(t) = 
\begin{cases}
2t\alpha & \text{for $0 \le t \le \tfrac{1}{2}$,}\\
\alpha + i(2t-1)\beta & \text{for $\tfrac{1}{2} \le t \le 1$.}
\end{cases}
\end{equation}
We need to show that $\zeta_j(t) = \Lambda(\gamma(t)) \zeta_j$ satisfies 
\begin{equation}
\|\zeta_j(t)\| < r, \quad \Im \zeta_j(t) \in V^- \quad
\text{for all $j = 1, \dots, n$ and $t \in [0, 1]$.}
\end{equation} 
In order to prove the first relation, we note that for any complex four vector $\zeta$, 
we have that $\| \zeta \|^2 = |\zeta^+|^2 + |\zeta^-|^2 + |\zeta^1|^2 + |\zeta^2|^2$, 
where $\zeta^\pm = (\zeta^0 \pm \zeta^3)/\sqrt{2}$.  For $t \in [0, \tfrac{1}{2}]$, this gives  
\begin{equation}
\|\zeta_j(t)\|^2 = |\e^{2t\alpha} \zeta_j^+|^2 + |\e^{-2t\alpha} \zeta_j^-|^2 + |\zeta_j^1|^2 + |\zeta_j^2|^2 \le 
\tfrac{1}{2}\|\zeta_j(0)\|^2 + \tfrac{1}{2}\|\zeta_j(\tfrac{1}{2})\|^2, 
\end{equation} 
by the convexity of the exponential function. For $t \in [\frac{1}{2}, 1]$, this gives
\begin{equation}
\|\zeta_j(t)\|^2 = |\e^{\alpha + i(2t-1)\beta} \zeta_j^+|^2 + |\e^{-\alpha-i(2t-1)\beta} \zeta_j^-|^2 + |\zeta_j^1|^2 + |\zeta_j^2|^2 
= \|\zeta_j(1)\|^2 < r^2, 
\end{equation}
It follows straightfowardly from these relations that $\|\zeta_j(t)\| < r$ for $t \in [0, 1]$ and all $j$.  
In order to prove the second relation, 
we note that, since $\Lambda(2t\alpha)$ are 
real resticed Lorentz transformations, we have that $\Im \zeta_j(t) = \Lambda(2t\alpha) \Im \zeta_j \in V^-$
for $t \in [0, \frac{1}{2}]$. To show that $\Im \zeta_j(t) \in V^-$ also for 
all $t \in [\tfrac{1}{2},1]$, it is sufficient show that $\Im \zeta_j(t)^\mu n_\mu > 0$ for any
real, future pointing timelike or null vector $n$, for all $j$. We have 
\begin{equation}
\Im \zeta_j(t)^\mu n_\mu = \sin \beta (2t-1) \, \Re \zeta_j^\mu n_\mu + \cos \beta (2t-1) \, \Im \zeta_j^\mu n_\mu, 
\end{equation}
which implies that
\begin{equation}
\sin \beta \, \Im \zeta_j(t)^\mu n_\mu = \sin \beta \tau \, \Im \zeta_j(1)^\mu n_\mu + \sin \beta(1-\tau) \, \Im \zeta_j(\tfrac{1}{2})^\mu n_\mu, 
\quad \tau = 2t-1. 
\end{equation}
The case $\beta = 0$ is trivial, and the case $\beta = \pi$ cannot occur, since otherwise we would have 
$\Im \zeta_j(1)^\mu n_\mu < 0$ for some future pointing timelike or null vector $n$, 
which cannot be since $\Im \zeta_j(1) \in V^-$ by assumption. We can therefore assume that $0 < \beta < \pi$. The above equation 
then displays $\Im \zeta_j(t)^\mu n_\mu$ as a positive linear combination of two positive numbers for $0 \le \tau \le 1$. 
This proves that $\Im \zeta_j(t) \in V^-$
for $\tfrac{1}{2} \le t \le 1$ and hence altoghether that $(\zeta_1(t), \dots, \zeta_n(t)) \in \mathcal{T}_n$ 
for all $0 \le t \le 1$. 
\end{proof}

Consider now a point $(\zeta_1, \dots, \zeta_n) \in \mathcal{T}_n$. Then we know that $\cW^{(J)}(\zeta_1, \dots, \zeta_n)$ transforms
according to the transformation law eq.~\eqref{lor'} for any rotation $R(\varphi)$ as in eq.~\eqref{Rdef}, and 
any complex Lorentz transformations $\Lambda(\alpha + i\beta)$. Applying this transformation rule in particular to 
the product $\Lambda(i\pi)R(\pi) = -1$, we find that 
\begin{equation}
\label{-1j}
\cW^{(J)}(-\zeta_1, \dots, -\zeta_n) = 
(-1)^j \cW^{(J)}(\zeta_1, \dots, \zeta_n)
\end{equation}
for all $(\zeta_1, \dots, \zeta_n) \in \mathcal{T}_n$, since $D(-1) = (-1)^{j}$. Moreover, since both sides of this 
equation are analytic functions, we find that this equation holds in fact for all
$(\zeta_1, \dots, \zeta_n)$ in the extended domain $\mathcal{T}_n'$. We will now use this 
this equation to get a relation between the $\cW^{(J)}$ for real arguments. We note however that
we cannot straightforwardly take the boundary value of both sides of the above equation 
as $\Im \zeta_j \to 0$ in order to get such a relation since $\Im \zeta_j$ has the opposite 
sign on both sides of the above equation and $\Im \zeta_j = 0$ can therefore not be approached
from within $V^-$ on both sides. 

To circumvent this problem, one considers special 
real points in $\mathcal{T}'_n$ defined as follows (compare~\cite[thm. 2.12]{sw}): Let
$n$ be the spacelike vector in $\mr^4$ given by $(0, 0, 0, 1)$, 
and consider the open, proper, convex and spacelike cone $\mathcal{K}$
in $\mr^4$ defined by the equation $\xi^\mu n_\mu > \|\xi\|$. 
\begin{lemma}
The extended domain $\mathcal{T}_n'$ includes the set 
\begin{equation}
\cJ_n = 
\{(\zeta_1, \dots, \zeta_n) \in \mr^{4n} \mid \| \zeta_j \| < r, \zeta_j \in \mathcal{K} \}, 
\end{equation}
and $\cJ_n$ is an open, real domain in $\mr^{4n}$. 
\end{lemma}
\begin{proof}
The last statement is obvious. 
We must show that if $(\xi_1, \dots, \xi_n) \in \cJ_n$, then 
there is $\alpha, \beta$ such that $(\Lambda(\alpha + i\beta) \xi_1, \dots, \Lambda(\alpha + i\beta) \xi_1)$ is 
in $\mathcal{T}_n$. One calculates that 
\begin{equation}
\Im \Lambda(i\beta) \xi_j = 
-\sin \beta \left(
\begin{matrix}
\xi_j^3\\
0\\
0\\
\xi_j^0
\end{matrix}
\right), \quad \|\Lambda(i\beta) \xi_j\| = \|\xi_j\|. 
\end{equation}
By definition, $\xi_j \in \mathcal{K}$ means that $\xi_j^\mu n_\mu > \|\xi_j\|$ and $\|\xi_j\| < r$, where 
$n = (0, 0, 0, 1)$. The first condition implies that $\xi^3_j > (|\xi^0_j|^2 + \dots + |\xi^3_j|^2)^{1/2} \ge |\xi^0_j|$, 
showing that $\Im \Lambda(i\beta) \xi_j \in V^-$ for all $j$ and any $0 < \beta < \pi$, and the second condition shows implies 
that $\|\Lambda(i\beta) \xi_j\| < r$ for all $j$. This proves the lemma. 
\end{proof}
One now notes that if $(\zeta_1, \dots, \zeta_n) \in \cJ_n$, and $(y_1, \dots, y_n) 
= f(\zeta_1, \dots, \zeta_n)$, then it follows that the difference vectors $y_j - y_k$ are 
all spacelike (and non-zero) since 
\begin{equation}
y_j - y_k = \sum_{j \le i < k} \zeta_i \in \mathcal{K}, 
\end{equation}  
by the convexity of $\mathcal{K}$. For such $(\zeta_1, \dots, \zeta_n) \in \mathcal{T}_n'$ 
we therefore know that eq.~\eqref{com'} holds. Combining this relation with eq.~\eqref{-1j}, we 
have therefore found that 
\begin{equation}
\label{WPCT''}
\cW^{(J)}(\zeta_1, \dots, \zeta_{n-1}, \zeta_n) = (-1)^j
\cW^{(\pi J)}(\zeta_{n-1}, \dots, \zeta_{1}, -\sum_{j=1}^{n} \zeta_j),  
\end{equation}
for all $(\zeta_1, \dots, \zeta_n) \in \cJ_n$. Moreover, 
since the set $\cJ_n$ forms an open real domain in the 
complex domain $\mathcal{T}'_n$, this equation will in fact hold for all $(\zeta_1, \dots, \zeta_n)
\in \mathcal{T}_n'$, so in particular for $(\zeta_1, \dots, \zeta_n) \in \mathcal{T}_n$. We now
take the boundary value of both sides of eq.~\eqref{WPCT''} as $\Im \zeta_j$ goes to zero while
keeping $\Im \zeta_j \in V^-$ for $1 \le j \le n-1$, which gives us
\begin{multline}
\bv_{\eta_j \to 0, \eta_j \in V^- \,\, \forall j \neq n}
\cW^{(J)}(\xi_1 + i\eta_1, \dots, \xi_{n-1} + i\eta_{n-1}, \xi_n + i\eta_{n}) = \\(-1)^j
\bv_{\eta_j \to 0, \eta_j \in V^- \,\, \forall j \neq n}
\cW^{(\pi J)}(\xi_{n-1} + i\eta_{n-1}, \dots, \xi_1 + i\eta_{1}, -\sum_{j=1}^n \xi_j + i\eta_{j}).  
\end{multline}
By eq.~\eqref{bv}, the left and right side of this equation are equal, in the distributional sense, to the left 
respectively right side 
of eq.~\eqref{WPCT'}. This proves 
the proposition and hence the theorem in the scalar, bosonic case. 

\bigskip

\paragraph{Proof of the theorem, general case:} The proof of the theorem when fields of arbitrary
spinor type and/or fermionic fields are present does not differ substantially from the scalar bosonic
case, so we will only briefly outline the main changes that occur in this more general case relative to 
the scalar bosonic case. 

If the functionals $\sigma^{(i)}$ in eq.~\eqref{key} carry abstract spinor indices collectively denoted 
$A_0$, and the fields $\phi^{(j_k)}$ carry spinor indices collectively denoted $A_k$, 
then the distributions $W_{\mu_1 \dots \mu_n}^{(J)}$ in eq.~\eqref{scexp} 
get replaced by spinor valued distributions,  
\begin{equation}
\bW_{\mu_1 \dots \mu_k \alpha_0 \dots \alpha_n}^{(J)}(y_1, \dots, y_n) \in \cD'(\times^n B_r), 
\quad 
\end{equation}
where $(J)$ stands for $(ij_1 \cdots j_n)$, and where $\alpha_j$ label the coordinate components corresponding 
the abstract spinor indices $A_j$ in a suitable trivialization of the spin bundle. 
By repeating the same kind of arguments as in the scalar, bosonic case (taking into account the 
definition of the map $\tilde I$, see appendix B), it is seen that the statement of 
the theorem in the general case can be reduced to the proof of the identity
\begin{equation}
\label{bwj'}
i^{F^{(i)}} (-1)^{M^{(i)}} \bW^{(J)}(y_1, \dots, y_{n}) 
= i^F (-1)^{M} (-1)^j \bW^{(\pi J)}(-y_n, \dots, -y_1).  
\end{equation}
The distributions $\bW^{(J)}$ now have the transformation behavior
\begin{equation}
\bW^{(J)}(\Lambda(L) y_1, \dots, \Lambda(L) y_n) = D^{(J)}(L)
\bW^{(J)}(y_1, \dots, y_n)
\end{equation}
for all $L \in {\rm SL}_2(\mc)$ with $\| L - 1\| < \delta$ 
and all $(y_1, \dots, y_n) \in \times^n B_{r-\delta}$, where $D^{(J)}(L)$ is now 
\begin{equation}
D^{(J)}(L)^{\beta_0 \dots \beta_n \nu_1 \dots \nu_j}_{\alpha_0 \dots \alpha_n\mu_1 \dots \mu_j} 
= D^{(i)}(L)^{\beta_0}_{\alpha_0} D^{(j_1)}(L)^{\beta_1}_{\alpha_1} 
\cdots D^{(j_n)}(L)^{\beta_n}_{\alpha_n} \Lambda(L)^{\nu_1}_{\mu_1} \cdots \Lambda(L)^{\nu_j}_{\mu_j},
\end{equation}
where $\Lambda(L)$ is the proper orthochronous Lorentz transformation corresponding to $L$ via the usual 
covering homomorphism ${\rm SL}_2(\mc) \to \cL$, and where $D^{(j)}(L)$ is the spinor representation of ${\rm SL}_2(\mc)$
corresponding to the spinor character of the field $\phi^{(j)}$. 
The commutation relations (iii) are modified to 
\begin{equation}
\bW^{(J)}(y_1, \dots, y_k, y_{k+1}, \dots, y_n) = 
\pm
\bW^{(\pi_{k,k+1} J)}
(y_1, \dots, y_{k+1}, y_k, \dots, y_n),
\end{equation}
whenever $y_k$ and $y_{k+1}$ are spacelike to each other with respect to the Minkowski 
metric $\eta_{\mu\nu}$, where $-$ is chosen if both fields $\phi^{(j_k)}$ and $\phi^{(j_{k+1})}$ are fermionic
and $+$ is chosen otherwise. (The permutation must also act in the obvious way on the spinor indices of 
$\bW^{(J)}$, but we have suppressed this.)  As in the scalar, bosonic case, the distributions
$\bw^{(J)} = f^* \bW^{(J)}$ are shown to be boundary values of analytic functions on $\mathcal{T}_n$. In 
order to pass to the extended tube, $\mathcal{T}_n'$, one considers the analytic 
family of transformations
\begin{equation}
L(\alpha + i\beta) = 
\left(
\begin{matrix}
\e^{i(\alpha + i\beta)/2} & 0 \\
0 & \e^{-i(\alpha + i\beta)/2} \\
\end{matrix}
\right), 
\end{equation}
in the group ${\rm SL}_2(\mc)$, so that the complex Lorentz-transformations 
$\Lambda(\alpha + i\beta)$ in eq.~\eqref{boost} are given by the pair $(L(\alpha
+ i\beta), L(\alpha - i\beta)) \in {\rm SL}_2(\mc) \times {\rm SL}_2(\mc)$
via the usual covering homomorphism ${\rm SL}_2(\mc) \times {\rm SL}_2(\mc) \owns
(L, M) \to \Lambda(L, M) \in {\mathcal L}_+(\mc)$. It is then straightforward
to prove the analog of lemma~\ref{extension} for the $\bw^{(J)}$.

Taking these modifications into account, one then proves in basically the same way as 
in the scalar, bosonic case that
\begin{equation}
\label{bwj}
\bW^{(J)}(y_1, \dots, y_{n}) 
= (-1)^{F(F-1)/2} (-1)^{m} (-1)^j \bW^{(\pi J)}(-y_n, \dots, -y_1) 
\end{equation}
for all $y_j \in B_r$ in the sense of distributions, where $F$ is 
the total number of fermion fields in the collection $\phi^{(j_1)}, \dots,
\phi^{(j_n)}$, and where $m = M^{(i)} +  M^{(j_1)} + \cdots  M^{(j_n)}$ 
is the total number of unprimed spinor indices represented 
by $\alpha_0 \dots \alpha_n$. One has
\begin{equation}
(-1)^{F(F-1)/2} = 
\begin{cases}
i^F & \quad \text{if $F$ is even,}\\
i^{F-1} & \quad \text{if $F$ is odd.}
\end{cases}
\end{equation}
If $F$ is even, then the field $\phi^{(i)}$ corresponding to the index $(i)$ in 
eq.~\eqref{key} has to be bosonic, $F^{(i)} = 0$, by remark (2) following
the theorem. If $F$ is odd, then $\phi^{(i)}$ has to be 
fermionic, $F^{(i)} = 1$. Hence in both cases, we get eq.~\eqref{bwj'}. 
This proves the theorem. 
\end{proof}

\paragraph{\bf Acknowledgements:} I would like to thank D.~Buchholz, K.~Fredenhagen, R.~Verch, 
and R.~M.~Wald for helpful and stimulating discussions. This work was supported by NSF grant
PHY00-90138 to the University of Chicago.


\appendix

\section{$\Sigma_A$ in Minkowski spacetime}

In this appendix, we show that when $\M = (\mr^4, \eta_{ab})$ is Minkowski spacetime and 
$\sigma$ is a linear functional on $\cA(\M)$ with bounded energy in the sense of eq.~\eqref{bde}, 
then $\sigma \in \Sigma_A(\M)$, i.e., the $n$-point functions of $\sigma$ satisfy the microlocal spectrum
condition~\eqref{wfc} in Minkowski spacetime. 
To see most clearly what is involved, we will give a formal argument, which can however
be made precise by the methods of~\cite[chap. 3]{sw}. 
Consider the distribution $u$ in $n+1$ spacetime variables defined formally by 
\begin{equation}
u(\xi_1, \dots, \xi_{n+1}) = \sigma(\e^{i\xi_1^\mu P_\mu} 
\phi^{(1)}(0) \e^{i\xi_2^\mu P_\mu} \cdots \e^{i\xi_n^\mu P_\mu} 
\phi^{(n)}(0) \e^{i\xi_{n+1}^\mu P_\mu}).
\end{equation}
The Fourier transform of $u$ is then (formally) given by 
\begin{equation}
\hat u(k_1, \dots, k_{n+1}) = \sigma(\delta(P - k_1) 
\phi^{(1)}(0) \delta(P-k_2) \cdots \delta(P - k_n) 
\phi^{(n)}(0) \delta(P-k_{n+1})).
\end{equation}
Since the spectrum of the energy-momentum operator $P$ lies entirely
within the forward lightcone $\bar V_+$, and since $\sigma$ has bounded 
energy in the sense of eq.~\eqref{bde}, it is easily seen that 
\begin{equation}
\supp \hat u \subset X_{p^0} \times (\times^{n-1} \bar V^+) \times X_{p^0}, 
\end{equation}
where $X_{p^0} = \{k \in \bar V^+ \mid k^0 \le p^0\}$. It follows 
by \cite[thm. 8.4.17]{h} that 
\begin{equation}
\WF_A(u) \subset (\times^{n+1} \mr^4) \times (\{0\} \times (\times^{n-1} \bar V^+) \times \{0\}).  
\end{equation} 
Now, it is easy to see from the transformation law $U(a) \phi^{(i)}(y) U(a)^* = \phi^{(i)}(y+a)$ 
of the local covariant fields under translations $y \to y + a$ that 
\begin{eqnarray}
\sigma(\phi^{(1)}(y_1) \cdots \phi^{(n)}(y_n)) &=& u(y_1, y_1 - y_2, \dots, y_{n-1} - y_n, y_n)\\
&=& f^* u(y_1, \dots, y_n),  
\end{eqnarray}
where the linear map $f: \mr^{4n} \to \mr^{4(n+1)}$ is defined by the last equation. 
Thus, by the transformation properties of the analytic wave front set under analytic maps, 
\begin{eqnarray}
&&\WF_A(\sigma(\prod_{k=1}^n \phi^{(k)}(y_k))) = f^* \WF_A(u)\\
&\subset& \{(y_1, k_1; \dots; y_n, k_n) 
\mid k_1 = p_{12}, k_n = -p_{(n-1)n}, k_i = p_{i(i+1)} - p_{(i-1)i}, \nonumber \\
&& \text{for $2 \le i \le n-1$, $p_{ij} \in \bar V^+$ for all $i,j$} \} \nonumber.
\end{eqnarray}
The right side of this inclusion is contained in the set $\Gamma_{\M}$ when $\M$ is 
Minkowski spacetime. This is seen by taking $G(p)$ in the definition~\eqref{gamtdef} of $\Gamma_\M$
to be the linear graph in which each point $x_i$ is connected to its predecessor $x_{i-1}$ by precisely one straight line, $e$,
decorated with the momentum $p_e = p_{(i-1)i} \in \bar V^+$. This proves that $\sigma \in \Sigma_A(\M)$.

\section{Spinors in curved spacetimes}

We here review the construction of spinors on a 4-dimensional curved spacetime $\M$, with a particular eye on 
the role played by the orientations. Our review follows closely~\cite{wbook}, to which we 
refer for details. Let $\M$ be 
a globally hyperbolic spacetime with space and time orientation $o = (T, \epsilon_{abcd})$. Let $x$ 
be a point in the spacetime. In the tangent space $T_x(\M)$ at $x$ we  
consider the set $F_x(\M)$ of all oriented, time oriented orthonormal frames 
$(e_\mu)^a$, (where $\mu = 0, 1, 2, 3$), meaning that 
\begin{equation}
g_{ab}(e_\mu)^a (e_\nu)^b = \eta_{\mu\nu}, \quad (e_0)^a \nabla_a T > 0, 
\quad \epsilon_{abcd} (e_0)^a(e_1)^b(e_2)^c(e_3)^d > 0. 
\end{equation}
Clearly, if $(e_\mu')^a$ is another such frame, 
then there is a unique proper orthochronous Lorentz transformation such that $(e'_\mu)^a = 
\Lambda^\nu_\mu (e_\nu)^a$. The frame bundle, $F(\M)$, is defined as the union of the 
spaces $F_x(\M)$ as $x$ runs over all points in $\M$. It has the stucture of a principal 
fibre bundle over $\M$, whose structure group is the proper orthochronous Lorentz group $\cL$ which 
acts upon elements in each fibre $F_x(\M)$ by transforming the orthonormal frames. Spinors over $\M$ can be defined if and
only if there exists a principal fibre bundle $S(\M)$, called ``spin-bundle'', 
with structure group ${\rm SL}_2(\mc)$ and base manifold $\M$ that covers the frame bundle 
in the sense that there is an onto map $f: S(\M) \to  
F(\M)$ such that the group action of ${\rm SL}_2(\mc)$ on the spin bundle 
corresponds to the group action of the Lorentz group on the frame bundle via 
the covering homomorphism ${\rm SL}_2(\mc) \to \cL$. A spin bundle need not exist
in a general curved spacetime, and if it exists, it need not be unique. The situation 
is however rather simple in the case when the spacetime $\M$ is simply connected, 
$\pi_1(\M) = 0$. In that case, a (necessarily unique) spin bundle will exist if and 
only if $\pi_1(F(\M)) = {\mathbb Z}_2$, and the spin 
bundle is in fact simply given by the the universal covering space of the frame bundle $F(\M)$.  
(Remember that the universal covering space $\tilde X$ of a topological space $X$ is the space of equivalence
classes of continuous paths $\gamma: [0,1] \to X$ with $\gamma(0) = x_0$, 
where two paths are equivalent if they can be composed to a closed path in 
$X$ that is homotopic to the trivial path given by $\gamma(\lambda) = x_0$ for all 
$\lambda \in [0,1]$.) 
Let $\chi: \N \to \M$ be an isometric, orientation and time orientation preserving embedding. 
If $(e_\mu)^a$ is an oriented and time-oriented orthonormal frame on $\N$, 
then clearly $\chi_* (e_\mu)^a$ will be such a frame over $\M$ and this defines
an embedding $\chi_*: F(\N) \to F(\M)$. This 
embedding lifts to a corresponding map between the covering spaces by defining 
its action on a path $\gamma: [0, 1] \to F(\N)$ to be the 
path $\gamma': [0, 1] \to F(\M)$ given by 
$\gamma'(\lambda) = \chi_* \gamma(\lambda)$ for all $\lambda \in [0, 1]$, because
it is easily seen that the equivalence class of $\gamma'$ only depends 
on the equivalence class of $\gamma$. Thus, if both $\N$ and 
$\M$ are simply connected and $\pi_1(F(\M)) = {\mathbb Z}_2$, 
then also $\pi_1(F(\N)) = {\mathbb Z}_2$, and we 
get a natural embedding map
\begin{equation}
\label{chi*}
\chi_*: S(\N) \to S(\M)
\end{equation}
which is compatible with the action of the group ${\rm SL}_2(\mc)$ 
on these spaces. (This can be seen by noting that, since ${\rm SL}_2(\mc)$
is the universal cover of $\cL$, every $L \in {\rm SL}_2(\mc)$ can be 
identified with an equivalence of continuous paths $\gamma_L: [0,1] \to \cL$ starting at
$1$ and ending at $\Lambda$, the element in $\cL$ covered by $L$.) 

If the spacetime $\M$ is {\em not} simply connected, i.e. $\pi_1(\M) = G \neq 0$, 
then one can show that a spin-bundle $S(\M)$ covering 
the frame bundle will exist if and only if the fundamental group of 
the frame bundle is isomorphic to a direct product
\begin{equation}
\label{pi1}
\psi: \pi_1(F(\M)) 
\cong {\mathbb Z}_2 \times G
\end{equation}
in the sense that every element of the form 
$(g_1, e_2)$ (with $e_2$ the identity element in $G$) corresponds to 
a path in $F(\M)$ that is homotopic to a path lying within a single fiber 
$F_x(\M)$, and that each path in $F(\M)$ 
corresponding to an element of the form $(e_1, g_2)$ (with $e_1$ the
identity in ${\mathbb Z}_2$)
projects down to a path in $\M$ that is homotopic to a path representing
$g_2 \in \pi_1(\M)$. In this case, one can define 
a spin bundle $S(\M)$ as the space of equivalence classes
of continuous paths in $F(\M)$, where two such paths are
now regarded as equivalent if their composition can be continuously
deformed to a path that corresponds to the group element of the form $(e_1, g_2)$ 
under the isomorphism $\psi$. Since this isomorphism 
is not necessarily unique\footnote{It is clear that 
the different possible choices for the isomorphism~\eqref{pi1} are 
in 1-to-1 correspondence with the non-unity automorphisms of the group
${\mathbb Z}_2 \times G$. These in turn are easily seen to be in 1-to-1 correspondence with the 
normal subgroups $H$ of $G$ such that $G/H = {\mathbb Z}_2$.}, there may now 
exist several inequivalent constructions of $S(\M)$, each 
corresponding to a particular choice for the isomorphism $\psi$.
A isometric embedding $\chi: \N \to \M$ between spacetimes admitting a spin structure, 
can therefore be lifted to a map $\chi_*$ as in eq.~\eqref{chi*} 
for one and only one choice of spin-structure over $\N$.  

Spinors in the spacetime $\M$ are constructed 
as elements in the vector bundles that are associated with the principal fibre bundle $S(\M)$. 
These are defined as follows. On the cartesian product $S(\M) \times \mc^2$
we define an equivalence relation $\sim$ by declaring two elements $(s, v)$ and $(s', v')$
to be equivalent if there is an element $L \in {\rm SL}_2(\mc)$ such that 
$s' = L^{-1} s$ and $v' = D(L)v$, where $L^{-1} s$ denotes the action of a group element $L$ on 
an element $s$ in the principal fibre bundle, and where $D$ is the fundamental 
representation of ${\rm SL}_2(\mc)$ on $\mc^2$. The space of equivalence classes
\begin{equation}
\cV(\M) = (S(\M) \times \mc^2)/\sim
\end{equation}
is then seen to be a vector bundle over $\M$ with each fibre isomorphic to 
$\mc^2$. Classical spinors fields over $\M$ (with an upper unprimed spinor index) 
are by definition sections in this vector bundle. 
This construction can be varied by replacing the space $\mc^2$ by the
dual space ${\mc^2}^*$ of complex linear functionals on $\mc^2$, or the space
$\bar \mc^{2*}$ of antilinear functionals on $\mc^2$, or the space $\bar \mc^2$ dual 
to $\bar \mc^{2 *}$, and by replacing the representation $D$ by the appropriate representations
of ${\rm SL}_2(\mc)$ on these spaces. We shall denote the corresponding 
vector bundles over $\M$ by $\cV^*(\M), \cV^{\prime *}(\M)$ and $\cV'(\M)$, respectively. 
They correspond to spinors with a lower unprimed, lower primed and upper primed index.
  
Suppose that we are given an isometric, orientation and causality preserving embedding, 
$\chi: \N \to \M$ between two oriented spacetimes $\N$ and $\M$ and suppose that 
each spacetime has a spin structure such that $\chi$ lifts to a map
$\chi_*$ as in eq.~\eqref{chi*} between the corresponding spin bundles.   
In this situation, we automatically get a map 
\begin{equation}
\label{chidef}
\chi_*: \cV(\N) \to \cV(\M), \quad [(s, v)]_\sim \to [(\chi_*(s), v)]_\sim 
\end{equation}
between the corresponding associated spin bundles (and likewise the bundles $\cV'(\M)$, $\cV^{\prime *}(\M)$ and
$\cV'(\M)$ as well as their tensor products).

\medskip

We finally explain the dependence of the above construction of spinors
on the choice of space and time orientation of the spacetime. 
Let $F(\M)$ be the bundle of frames that are orthogonal
with respect to the metric $g_{ab}$ and that are oriented and 
time oriented with respect to a time and space orientation $o = (T, \epsilon_{abcd})$
on $\M$, and let $F(\overline \M)$ be the bundle of orthonormal frames that 
are oriented with respect to the opposite time and space orientation, $-o=(-T, \epsilon_{abcd})$. 
Then these bundles are naturally isomorphic under the map
\begin{equation}
I: F(\M) \to F(\overline \M), \quad (e_\mu)^a \to (-e_\mu)^a,  
\end{equation}
since a tetrad $(e_\mu)^a$ is positively oriented with respect to $o$ if and only if the 
tetrad $(-e_\mu)^a$ is positively oriented with respect to $-o$.
As explained above, a construction of a spinor bundle $S(\M)$ covering the 
frame bundle $F(\M)$ is equivalent to a choice of isomorphism $\psi: \pi_1(F(\M)) \to
{\mathbb Z}_2 \times G$, and a construction of a spinor bundle $S(\overline \M)$ in the 
spacetime $\overline \M$ with the opposite orientations 
is likewise equivalent to a choice of isomorphism $\psi': \pi_1(F(\overline \M)) \to 
{\mathbb Z}_2 \times G$. It is possible to see that the map $I$ will lift to 
a corresponding bundle isomorphism $\tilde I$ between the spinor bundles $S(\M)$
and $S(\overline \M)$ if and only if $\psi$ and $\psi'$ are compatible in the sense that
$\psi' \circ I \circ \psi^{-1}$ is the identity homomorphism in 
$\pi_1(\M) \times {\mathbb Z}_2$. By changing $\psi$ or $\psi'$ if 
necessary, we can therefore always assume that there is indeed a natural map $\tilde I$
identifying the spin bundles $S(\overline \M)$ and $S(\M)$. From the above constructions 
it is then clear that this map will induce a corresponding map
\begin{equation}
\tilde I: \cV(\M) \to \cV(\overline \M)
\end{equation}
between the corresponding associated vector bundles of which the spinors are elements, and 
similar statements hold for the bundles $\cV'(\M), \cV^*(\M), \cV^{\prime *}(\M)$, as well 
as their tensor products.
This provides us with a natural identification of spinors over the spacetimes $\M$ and $\overline \M$.

\section{Proof of lemma~\ref{appclem}}

For the convenience of the reader, we repeat the statement of the lemma:
\begin{lemma}
Let $u^{(s)} \in \cD'(X)$, $X$ an open subset of $\mr^m$, a 
family of distributions that depends analytically on $s \in I = (a, b)$ with respect to 
conic sets $K^{(s)} \subset X \times (C \setminus \{0\})$, where $C$ is 
a closed, proper convex cone in $\mr^m$. Suppose that
\begin{equation}
\frac{d^k}{d s^k} u^{(s)} |_{s=s_0} = 0
\end{equation} 
for all $k$ and some $s_0 \in I$. Then $u^{(s)} = 0$ for all $s \in I$. 
\end{lemma}
\begin{proof}
Let us set $\tilde X = I \times X, \tilde x = (s, x)$, and view the family of distributions $u^{(s)}$ as defining 
a distribution $\tilde u \in \cD'(\tilde X)$. Since the set $C$ is proper, closed and
convex, we may without loss of generality assume that $C = \{k \in \mr^m \mid k \cdot n \ge \delta 
\|n\| \|k\| \}$, where ``dot'' is the standard inner product for vectors in $\mr^m$, 
where $\| \, \cdot \, \|$ is the corresponding norm, where $0 < \delta < 1$ and where 
$n$ is some nonzero vector in $\mr^m$. Let $\tilde n = (0, n) \in \mr^{m+1}$ and 
consider the quantity
\begin{equation}
c = \inf_{(\tilde x, \tilde k) \in \WF_A(\tilde u) \restriction \tilde X_0} 
\frac{\tilde k \cdot \tilde n}{\|\tilde k\| \|\tilde n\|}, 
\end{equation} 
where $\tilde X_0$ is any closed compact subset of $\tilde X$.
We claim that $c > 0$. Since the analytic wave front is closed the infimum 
is achieved for some $(\tilde x_0, \tilde k_0) \in \WF_A(\tilde u) \restriction \tilde X_0$. If $c \le 0$, 
we have consequently $0 \ge \tilde k_0 \cdot \tilde n = k_0 \cdot n$, 
\begin{equation}
k_0 = {^t f}^{(s)\prime}(x_0) \tilde k_0, \quad \tilde x_0 = f^{(s)}(x_0)
\end{equation}
where $f^{(s)}: X \to \tilde X$ is the embedding map. It is therefore
not possible that $k_0 \in C$, unless $k_0 = 0$. This is however in contradiction 
with the assumption of the lemma, since the analyticity of $u^{(s)}$ with respect to $s$ implies that when 
$(\tilde x_0, \tilde k_0) \in \WF_A(\tilde u)$, then necessarily $k_0 \in C \setminus \{0\}$. 
We must therefore have that $c > 0$, and consequently that
\begin{equation}
\WF_A(\tilde u) \restriction \tilde X_0 \subset \tilde X_0 \times \tilde C, \quad 
\tilde C = \{\tilde k \in \mr^{m+1} \mid \tilde k \cdot \tilde n \ge c 
\|\tilde n\| \|\tilde k\| \}, 
\end{equation}
and we may assume without loss of generality that $c < 1$. The cone $\tilde C$ is the dual of the open cone 
consisting of all $\tilde x \in \mr^{m+1}$, such that $\tilde x \cdot \tilde n > (1-c) \|\tilde n\| \|\tilde x\|$. 
By thm.~\ref{bvthm}, we can therefore conclude that there is a function $\tilde U$ that is analytic in the complex domain 
consisting of all $\tilde x + i \tilde y \in \mc^{m+1}$
for which $\tilde y \cdot \tilde n > (1-c) \| \tilde n \| \| \tilde y \|$, and  
$\tilde x \in \tilde X_0$ so that $\tilde u$ is the boundary value of $\tilde U$ as $\tilde y \to 0$,  
\begin{equation}
\label{bv'''}
\tilde u(s, x) = \bv_{(t,y) \to 0, n \cdot y > (1-c)\|n\| \sqrt{\|y\|^2 + t^2}} 
\tilde U(s + it, x + iy),  
\end{equation}
where we are now writing $\tilde x = (s, x), \tilde y = (t, y)$, and where we have used the 
definition $\tilde n = (0, n)$. We may set $t = 0$ on the right side of this equation when $y \cdot n > 
(1-c)\|n\|\|y\|$, and take $k$ derivatives with respect to $s$ of both sides of the equation. Setting
$s = s_0$ and using the assumption of the lemma, this gives
\begin{equation}
0 = \bv_{y \to 0, n \cdot y > (1-c)\|n\|\|y\|} 
\frac{d^k}{ds^k} \, \tilde U(s_0, x + iy) \quad \forall k.
\end{equation}
We already know that the function $x + iy \to d^k/ds^k \,\tilde U(s_0, x + iy)$ is analytic when 
$n \cdot y > (1-c)\|n\|\|y\|$, and we have now found that its distributional boundary values as $y \to 0$
vanish. We therefore conclude, by the ``edge-of-the-wedge theorem'' (see e.g.~\cite[thm. 2.17]{sw}), 
that this function itself has to vanish. Therefore 
\begin{equation}
\tilde U(s, x + iy) = \sum_{k=0}^\infty \frac{(s-s_0)^k}{k!} \frac{d^k}{ds^k} \tilde U(s_0, x + iy) = 0
\end{equation}
for sufficiently small $|s-s_0|$, and hence for all $s$ in $I$. Thus, by eq.~\eqref{bv'''}, $\tilde u(s, x) = u^{(s)}(x) = 0$ 
in the sense of distributions for all $s \in I$. 
\end{proof}


\begin{thebibliography}{99}

\bibitem{w} K. Wilson, ``Non-Lagrangian Models of Current Algebras,'' Phys. Rev. {\bf 179}, 
1499 (1969); ``Anomalous Dimensions and the Breakdown of Scale Invariance in 
Perturbation theory,'' Phys. Rev. {\bf D2}(8), 1478-1493 (1970)

\bibitem{z} 
W.~Zimmermann,
``Normal Products And The Short Distance Expansion In The Perturbation Theory Of Renormalizable Interactions,''
Annals Phys.\  {\bf 77}, 570 (1973)
[Lect.\ Notes Phys.\  {\bf 558}, 278 (2000)]

\bibitem{wz}
K. Wilson and W. Zimmermann, ``Operator Product Expansions and Composite Field
Operators in the General Framework of Quantum Field Theory,'' Commun. Math. Phys.
{\bf 24}, 87-106 (1972)

\bibitem{fj}
K. Fredenhagen and M. J\"orss, ``Conformal Haag-Kastler Nets, Pointlike
Localized Fields and the Existence of Operator Product Expansions,''
Commun. Math. Phys. {\bf 176}, 541-554 (1996)

\bibitem{ss} S. Schlieder and E. Seiler, ``Remarks Concerning the Connection
between Properties of the 4-Point Function and the Wilson-Zimmermann Expansion,''
Commun. Math. Phys {\bf 31}, 137-159 (1973)

\bibitem{l}
M.~Luscher,
``Operator Product Expansions On The Vacuum In Conformal Quantum Field Theory In Two Space-Time Dimensions,''
Commun.\ Math.\ Phys.\  {\bf 50}, 23 (1976).

\bibitem{b} 
H. Bostelmann, ``Lokale Algebren und Operatorprodukte am Punkt,'' Doktorarbeit, Universit\"at 
G\"ottingen (2000), available at {\tt http://www.lqp.uni-goettingen.de/papers/00/12/00121700.html}

\bibitem{fh}
K. Fredenhagen, J. Hertel, ``Local Algebras of Observables and Pointlike
Localized Fields,'' Commun. Math. Phys. {\bf 80}, 555-561 (1986)

\bibitem{bw}
D.~Buchholz and E.~H.~Wichmann,
``Causal Independence And The Energy Level Density Of States In Local Quantum Field Theory,''
Commun.\ Math.\ Phys.\  {\bf 106}, 321 (1986).

\bibitem{bfk}
R. Brunetti, K. Fredenhagen and M. K\"ohler, ``The microlocal spectrum 
condition and Wick polynomials on curved spacetimes,'' Commun. Math.
Phys. {\bf 180}, 633-652 (1996) [arXiv:math-ph/9903028]

\bibitem{hw3} S. Hollands and R. M. Wald: Work in progress

\bibitem{sw} 
R. F. Streater and A. S. Wightman, ``PCT, Spin and Statistics and All That,''
Benjamin, New York (1964)

\bibitem{pj}
W. Pauli, ``Exclusion Principle, Lorentz Group and Reflection of Space-Time and
Charge,'' p. 30 in: ``Niels Bohr and the Devolopment of Physics,'' W. Pauli (ed.), 
Pergamon Press, New York (1955); 
R. Jost, ``Eine Bemerkung zum CPT-Theorem,'' Helv. Phys. Acta {\bf 30}, 409 (1957)

\bibitem{bo}
H.~J.~Borchers and J.~Yngvason,
``On the PCT-theorem in the theory of local observables,''
[arXiv:math-ph/0012020.]

\bibitem{bfv} R. Brunetti, K. Fredenhagen and R. Verch:
``The generally covariant locality principle: A new paradigm for local  quantum physics,''
[arXiv:math-ph/0112041].

\bibitem{hw1} 
S.~Hollands and R.~M.~Wald,
``Local Wick polynomials and time ordered products of quantum fields in  curved spacetime,''
Commun.\ Math.\ Phys.\  {\bf 223}, 289 (2001)
[arXiv:gr-qc/0103074]

\bibitem{hw2} 
S.~Hollands and R.~M.~Wald,
``Existence of local covariant time ordered products of quantum fields in  curved spacetime,''
Commun. Math. Phys. (in print)
[arXiv:gr-qc/0111108]

\bibitem{haag}
R. Haag: ``Local Quantum Physics,'' Springer Verlag, Berlin (1992)

\bibitem{ver} 
R.~Verch, 
``A spin-statistics theorem for quantum fields on curved spacetime  manifolds in a generally covariant framework,''
Commun.\ Math.\ Phys.\  {\bf 223}, 261 (2001)
[arXiv:math-ph/0102035].

\bibitem{h}
L. H\"ormander, ``The Analysis of Linear Partial Differential Operators I,''
Springer Verlag, Berlin 1983

\bibitem{kw} 
B. S. Kay and R. M. Wald: ``Theorems on the Uniqueness
and Thermal Properties of Stationary, Nonsingular, Quasifree
States on Spacetimes with a Bifurcate Killing Horizon,''
Phys. Rep. {\bf 207} 49--136 (1995)

\bibitem{rad} 
M.~J.~Radzikowski,
``Micro-Local Approach To The Hadamard Condition In Quantum Field Theory On Curved Space-Time,''
Commun.\ Math.\ Phys.\  {\bf 179}, 529 (1996).

\bibitem{ver1}
A.~Strohmaier, R.~Verch and M.~Wollenberg,
``Microlocal analysis of quantum fields on curved spacetimes: Analytic  wavefront sets and Reeh-Schlieder theorems,''
[arXiv:math-ph/0202003].

\bibitem{st}
O. Steinmann, ``Perturbation Expansions in Axiomatic Field Theory,'' Lect. Notes in Phys. {\bf 11}, Springer
Verlag, Berlin (1971)

\bibitem{wbook} R. M. Wald, ``General Relativity,'' Unversity of Chicago Press, Chicago (1984)

\end{thebibliography}
\end{document}